\documentclass[aip,jcp,superscriptaddress,amsfonts,graphicx,preprint]{revtex4-2}
\usepackage[utf8]{inputenc}
\usepackage[T1]{fontenc}
\usepackage{tgtermes}
\usepackage{amsmath, amssymb}
\usepackage{mathtools}
\usepackage{breqn}
\usepackage{graphicx}
\usepackage{float}
\usepackage{bm}
\usepackage{siunitx}
\usepackage{physics}
\AtBeginDocument{\RenewCommandCopy\qty\SI}
\usepackage{verbatim}
\usepackage{enumitem}
\usepackage{booktabs}
\graphicspath{{fig/}}
\usepackage{hyperref}
\usepackage{multirow,xcolor}
\hypersetup{
    colorlinks = true,
    linkcolor = blue,
    linkbordercolor = white,
    citecolor=blue,
    urlcolor=black
}
\usepackage[export]{adjustbox}
\usepackage{subfig,caption}
\newcommand{\bfr}{\mathbf{r}}
\newcommand{\bfv}{\mathbf{v}}
\newcommand{\Rb}{\mathbf{R}}\newcommand{\mR}{\mathcal{R}}

\newcommand{\Pb}{\mathbf{P}}

\newcommand{\cev}[1]

\makeatletter
\let\cat@comma@active\@empty
\makeatother

\begin{document}

\title{Surface Hopping, Electron Translation Factors, Electron Rotation Factors, Momentum Conservation, and Size Consistency}
\date{\today}

\author{Vishikh Athavale}
\affiliation{Department of Chemistry, University of Pennsylvania, Philadelphia, Pennsylvania 19104, USA}
\author{Xuezhi Bian}
\affiliation{Department of Chemistry, University of Pennsylvania, Philadelphia, Pennsylvania 19104, USA}
\author{Zhen Tao}
\affiliation{Department of Chemistry, University of Pennsylvania, Philadelphia, Pennsylvania 19104, USA}
\author{Yanze Wu}
\affiliation{Department of Chemistry, University of Pennsylvania, Philadelphia, Pennsylvania 19104, USA}
\author{Tian Qiu}
\affiliation{Department of Chemistry, University of Pennsylvania, Philadelphia, Pennsylvania 19104, USA}
\author{Jonathan Rawlinson}
\affiliation{Department of Mathematics, University of Manchester, Manchester M13 9PL, UK}
\author{Robert G. Littlejohn}
\affiliation{Department of Physics, University of California, Berkeley, California 94720, USA}
\author{Joseph E. Subotnik}
\email{subotnik@sas.upenn.edu}
\affiliation{Department of Chemistry, University of Pennsylvania, Philadelphia, Pennsylvania 19104, USA}

\begin{abstract}
For a system without spin-orbit coupling, the (i) nuclear plus electronic linear momentum and (ii) nuclear plus orbital electronic angular momentum are good quantum numbers. Thus, when a molecular system undergoes a nonadiabatic transition, there should be no change in the total linear or angular momentum. Now, the standard surface hopping algorithm  ignores the electronic momentum and indirectly equates the momentum of the nuclear degrees of freedom to the total momentum. However, even with this simplification, the algorithm still  does {\em not} conserve either the nuclear linear or the nuclear angular momenta. Here, we show that one way to address these failures is to dress the derivative couplings (i.e. the hopping directions) in two ways: (i) we disallow changes in the nuclear linear momentum by working in a translating basis (which is well known and leads to electron translation factors [ETFs]); and (ii) we disallow changes in the nuclear angular momentum by working in a basis that rotates around the center of mass (which is not well-known and leads to a novel, rotationally removable component of the derivative coupling that we will call electron rotation factors [ERFs] below, cf. Eq. \ref{easy1final}).
 The present findings should be helpful in the short term as far as interpreting surface hopping calculations for singlet systems (without spin) and then developing new surface hopping algorithm in the long term for systems where one cannot ignore the electronic orbital and/or spin angular momentum. 
\end{abstract}

\maketitle

\section{Introduction} 

Surface hopping is perhaps the most popular semiclassical nonadiabatic dynamics algorithm\cite{tully:fssh,barbatti:2011:review}. The idea behind surface hopping is very simple: one propagates dynamics on one adiabatic surface for some period of time, and one hops between surfaces so as to account for nonadiabatic transitions. The algorithm is routinely used today\cite{tretiak:2014:acr} because of it simplicity, its low cost, and its applicability to {\em ab initio} problems. One of the algorithm's main selling points is the ability to more or less simulate the correct quasi-classical equilibrium distribution with detailed balance\cite{tully:2005:detailedbalance,tully:2008:detailedbalance}. {This feat is achieved because the algorithm conserves energy (and does not allow frustrated hops).}


Interestingly, however, standard surface hopping does {\em not} conserve linear momentum or angular momentum in the following sense.  Consider the case of one H atom (i.e one electron and one nucleus) with the electron initially in a $3p_x$  orbital. If ${\bf r}$ represents the electronic position and ${\bf R}$ the nuclear position, this initial wavefunction can be represented as:
\begin{align}
    \Psi_i({\bf r},{\bf R}) = \chi({\bf R}) \phi_{3p_x}({
    \bf r};{\bf R})
\end{align}
In Cartesian coordinates, the derivative coupling between a $3p_x$ and a $1s$ state is nonzero\cite{furche:2010:nact,kutzelnigg:2007:mp}. Thus, a naive reading of Eqs. \ref{hop2} and \ref{rescale} below would suggest that if an H atom is traveling in the $x$-direction, there is a finite chance that the electron will relax to the $1s$ state and the atom will accelerate from $\Pb_i$ to $\Pb_f$ (and gain speed) in the $x$-direction.  The final wavefunction would be represented as:
\begin{align}
        \Psi_f({\bf r},{\bf R}) = \chi'({\bf R}) \phi_{1s}({\bf r};{\bf R}) 
\end{align}

Unfortunately,  this hypothetical scenario makes clear the surface hopping cannot maintain momentum conservation.  For the experiment just described,
a  naive quantum calculation for both initial wavefunctions yields: 
\begin{align}
    \left<\Psi_i \middle | \hat{p}_e \middle| \Psi_i \right>  = 
        \left<\Psi_f \middle | \hat{p}_e \middle| \Psi_f \right>  = 0
\end{align}
Thus, according to the most naive interpretation of the Born-Oppenheimer approximation, both the initial and final electronic states have zero momentum {(which was recognized long ago\cite{patchkovskii:2012:jcp:electronic_current})}. And so, if the nucleus were to change its momentum (from $\Pb_i$ to $\Pb_f$), the linear momentum would not be conserved. On both mathematical and physical grounds, it is clear that if the electronic momentum can be ignored, the linear momentum of the nuclear degrees of freedom (which is then equal to the total molecular momentum) cannot induce a nonadiabatic transition. 
Over the last few decades, a few authors have  {studied realistic molecules  with realistic derivative couplings (as opposed to Tully-style models) and} discussed the fact that standard surface hopping fails to conserve linear momentum\cite{wardlaw:1998:momentum_sh,drukker:1999:review_fssh,fatehi:2011:dercouple, fatehi:2012:dercouple,truhlar:2020:project_rot_nac}.



Now, the thought experiment above lays bare some of the subtleties associated with  electronic momentum within the confines of the Born-Oppenheimer approximation  \cite{mead_moscowitz:1967:dipole_vs_length,nafie:1983:jcp:elcurrent}, and then demonstrates that those subtleties can erupt into big problems for a surface hopping simulation.
Within the scenario imagined above, one must wonder: why is the initial electronic momentum zero? After all, is not the electron being dragged with the nucleus so that, if the nucleus initially has velocity $v$, must not also the electron?  These are crucial questions because one cannot equate the {\em molecular} (nuclear + electronic) center of mass (which is a good quantum number) with the momentum of the {\em nuclear} center of mass  (which is not a good quantum number). Indeed, there has been a great deal of work on identifying and quantifying the current within a Born-Oppenheimer framework\cite{patchkovskii:2012:jcp:electronic_current} and the bottom line is that one must allow for state crossings (and derivative couplings) if one seeks to capture a nonzero electronic momentum.

Alas, even though standard Tully-style surface hopping aims to  go beyond the Born-Oppenheimer approximation and explicitly capture the entanglement of nuclear motion with electronic transitions through derivative couplings, the algorithm effectively ignores electronic momentum and does not address the questions hypothesized above.  
This omission would appear reasonable on several grounds. First, on physical grounds, one can wonder whether including electronic momentum is actually important when one traverses a conical intersection? Second, on practical grounds, like all semiclassical nonadiabatic dynamics algorithms, the surface hopping algorithm treats electrons and nuclei differently and so understanding and comparing electronic versus nuclear momentum is already  tricky. Indeed,  there is a large literature discussing different ways that electronic observables can be calculated within a surface hopping ansatz\cite{stock:1997:surfacehop,markland:2013:surfacehop_marcus,shs:2010:pcet_fssh, shs:2014:marcustheory_fssh,landry:2011:marcus_fssh,vbk:2010:furan},  and until recently\cite{subotnik:2013:qcle_fssh_derive,kapral:2016:chemphys_fssh}, it was difficult to assess the origins of Tully's surface hopping algorithm\cite{landry:2013:electronicproperties} in the first place. 
For these reasons, many practitioners in the nonadiabatic dynamics community have largely ignored electronic momentum.  That being said, practical solutions to nuclear momentum conservation have emerged over the past decade. The bottom line is that, if one works in a translating (as opposed to fixed) basis of electronic orbitals, one can derive reasonable electron translation factors\cite{fatehi:2011:dercouple} that are appropriate for rescaling momentum and momentum conservation (see below).

We now come to the key question motivating the present paper. Unlike the case of linear momentum, the subject of angular momentum conservation is not widely discussed in the literature in the context of semiclassical nonadiabatic dynamics. And so, after resolving that the total nuclear momentum cannot induce a change in electronic state, one must ask:
Can the total angular momentum of the nuclei induce such  a change? And if not, how should we modify the surface hopping algorithm (which will allow such transitions)? Without thinking too hard, one might presume that angular momentum could induce nonadiabatic transitions because we know that angular momentum cannot be fully disentangled from vibrational motion (which leads to the Coriolis force), and the derivative coupling in an angular direction is certainly nonzero\cite{yarkony:1989:jcp_emailme_ang}.  And yet,  if the electronic states carry zero angular momentum, the analogous argument about angular momentum must dictate that,  
just as for the case of linear momentum,  the nuclear angular momentum should be conserved and should not induce nonadiabatic electronic transitions between electronic eigenstates.  For this reason, \citeauthor{truhlar:2020:project_rot_nac} have suggested\cite{truhlar:2020:project_rot_nac} 
that, upon rescaling momentum after a hop, one should project the new momentum to make sure that one has not changed the total angular momentum.
That being said, while a naive projection ( as in Eq. \ref{eq-truhlar} below) is {\em ad hoc}, the question remains: is there a more general (and physical) means by which one can derive and insist that the overall nuclear angular momentum not dictate nonadiabatic transitions? In particular, given that moving to a translating basis is enough to find a momentum rescaling direction that conserves linear momentum, is there an analogous approach of moving to a rotating basis whereby one can remove the contribution of overall angular momentum from the derivative coupling? In this paper, we will provide a partial answer to this problem and offer a simple means for removing {\em both} the rotating and translating components of the derivative coupling at the same time.

Before concluding this introduction, an important caveat is in order. The discussion above was predicated on our being able to isolate eigenstates of the electronic Hamiltonian with formally zero electronic linear and/or angular momentum.  Mathematically, this means that  we must restrict ourselves to the case of the standard non-relativistic electronic Hamiltonian, where there is time-reversal symmetry but there are no spin-orbit  couplings or fine structure. 
As discussed above, in such a case,  within the context of surface hopping,
the total momentum of a trajectory moving along a given adiabatic surface is unambiguously equal to the nuclear momentum and therefore,  for a hop from electronic state $J$ to electronic state $K$, 
one must  conserve the {\em nuclear} linear and angular momentum: we will offer a physically-motivated means of achieving such nuclear momentum conservation below, in Eq. \ref{easy1final}. Nevertheless, even if we can figure out how to conserve the {\em nuclear} momentum, subtle questions still remain regarding how best to treat {\em electronic} linear and angular momentum when they cannot be ignored; we will discuss this problem further in Sec. \ref{sec-conc} and offer one alternative solution in the Appendix.



An outline of this paper is as follows. In Sec. \ref{sec-rev}, we will begin our analysis by quickly reviewing the surface hopping algorithm and we will remind the reader of why one would normally expect that the surface hopping algorithm should obey linear and angular momentum conservation according to Noether's theorem\cite{goldstein}.  Thereafter, in Sec. \ref{sec-lin}, we will dig further into the algorithm and show specifically why the algorithm fails to conserve linear momentum, a topic of much interest over the last decade, and we will review the concept of electron translation factors (ETFs).  In Sec. \ref{sec-ang}, we will extend the discussion from Sec. \ref{sec-lin} to the case of angular momentum, and we provide a new expression for a derivative coupling in a rotating frame (leading to the concept of eletronic rotational factors [ERFs]).  In Sec. \ref{sec-res}, a few computational results are presented. In Sec. \ref{sec-conc}, we conclude and discuss future directions, especially the future possibilities for understanding the coupling between nuclear motion and electronic spin.

Henceforward, we will use the following notation. We let
$\alpha, \beta, \gamma, \delta, \kappa$ index directions in three-dimensional space ($\mathbb{R}^3$). All vectors in $\mathbb{R}^3$ are in bold, while all operators have a hat.  The labels $J,K$ index adiabatic electronic states, while the labels $A,B,C$ index nuclear positions: $\Rb_A, \Rb_B,\Rb_C$. We let $N$ denote the number of nuclei. We denote vectors in $N$ (or $3N$) dimensional space with an overhead arrow, e.g. $\vec{R} = \left(\Rb_A, \Rb_B, \cdots, \Rb_N \right)$.

\section{Review of Surface Hopping and Noether's Approach to Momentum Conservation} \label{sec-rev}

Before we present our findings, it will be helpful to review the relevant surface hopping equations and the standard Noether approach to linear and angular momentum conservation.

\subsection{Overview of Surface Hopping}

Within a surface hopping algorithm, one propagates a trajectory along an adiabatic path and one integrates the electronic density matrix:
\begin{align}
    \dot{\rho}_{JK} = \sum_{L} \frac{-i}{\hbar}\left( H_{JL} \rho_{LK} -  \rho_{JL}H_{LK} \right)
\end{align}
Here, $\hat{H}$ is the electronic Hamiltonian, 
\begin{align}
\label{hel}
    \hat{H} = \sum_i \frac{\hat{\bf p}_i\cdot \hat{\bf p}_i}{2m_e} + V\left(\hat{\bf r}_1, \cdots, \hat{
    \bf r}_n, \Rb_1, \ldots, \Rb_N\right)
\end{align}
Let $J$ be the active surface. To calculate a hopping rate from surface $J$ to surface $K$, one decomposes the rate of change of population, $\dot{\rho}_{JK}$.  From this analysis, one finds a  very simple hopping probability of the form:
\begin{align}
\mbox{Prob}_{J \rightarrow K} = \Theta(\Gamma_{J \rightarrow K})
\end{align}
where $\Theta$ is the Heaviside function 
\begin{align}
    \Theta(x) = \left\{ 
    \begin{array}{c}
    x \; \; \; \; x > 0 \\
    0 \; \; \; \; x < 0
    \end{array}
    \right.
\end{align}

and 
\begin{align}
\label{hop}
    \Gamma_{J\rightarrow K} = 2\Re \left( \frac{ \left[ \frac{i}{\hbar}H_{JK} + \sum_{B \alpha} d^{B\alpha}_{JK}  v_{B \alpha} \right]\rho_{KJ}}{\rho_{JJ}}\right).
\end{align}

Here $v_{B \alpha}$ is the classical velocity of nucleus  $B$ in the $\alpha$ direction.  In practice, as stated above, one almost always uses an adiabatic basis,
so that $H_{JK}= 0$ and all hopping comes from the second term above:
\begin{align}
\label{hop2}
\Gamma_{J\rightarrow K} = 2 \mbox{Re} \left(\frac{ \sum_{B \alpha} d^{B\alpha}_{JK}  v_{B \alpha} \rho_{KJ}}{\rho_{JJ}}\right)
\end{align}
Here,  $d^{B \alpha}$ is the derivative coupling as caused by moving nucleus $B$ in the $\alpha$ direction:
\begin{align}
\label{hellman}
    d_{JK}^{B\alpha} = \frac{\left<\Phi_J \middle| \frac{\partial \hat{H}}{\partial R_{B \alpha}} \middle| \Phi_K \right>}{E_K -E_J}
\end{align}

After a hop from state $J$ to state $K$, semiclassical theory dictates that one rescale the nuclear momentum in the direction of the derivative coupling $({\bf d}_{JK})$:

\begin{align}
\label{rescale}
    \vec{P} \rightarrow  \vec{P} +   \epsilon \vec{d}
    \end{align}

The magnitude of the jump is dictated by solving for $\epsilon$ after mandating energy conservation:
\begin{align}
\label{rescale_mag}
    \left(\vec{P} +   \epsilon \vec{d} \right) \cdot \frac{1}{M} \cdot \left(\vec{P} +   \epsilon \vec{d} \right) + V_K = 
        \vec{P} \cdot \frac{1}{M} \cdot \vec{P} + V_J 
\end{align}

\subsection{Conservation of Linear and Angular Momentum in Principle}

If we want the surface hopping algorithm  to conserve linear and angular nuclear momenta, the key question is whether or not a surface hop in the direction of the derivative coupling will preserve the linear and angular momentum of the nuclei.
Thus, the form of the derivative coupling emerges as the fundamental quantity of interesting in surface hopping.  Note that the Hellman-Feynman expression in Eq. \ref{hellman} is not entirely equivalent to a gradient, but for the purposes of understanding the relevant conservation laws, it will be helpful to imagine that we can express the derivative coupling as the gradient of some abstract function $X$ (which is a surrogate for the Hamiltonian):

\begin{align}
\label{xderv}
    d_{JK}^{A \alpha} \stackrel{?}{=} \frac{\partial X_{JK}}{\partial R_{A \alpha}}
\end{align}

At this point, consider first the case of linear momentum.  If we wish to conserve the linear momentum during the course of a hop,  the derivative couplings must satisfy:
\begin{align}
\label{zero}
    \sum_A d_{JK}^{A\alpha} =  0    
\end{align}
In principle, using Noether's theorem, one can derive Eq. \ref{zero} using the translational invariance of the matrix ${\bf X}$.
More precisely, let $\mathbf{\Delta}$ be a vector in 3 dimensions, and let us imagine translating every nucleus by $\mathbf{\Delta}$. Let $\vec{\Delta} = \left( \mathbf{\Delta}, \mathbf{\Delta}, \ldots, \mathbf{\Delta}\right)$  be the corresponding $3N$ dimensional translation vector, and  let $\vec{R} = \left( {\bf R}_1, \ldots, {\bf R}_N \right)$ be the corresponding $3N$ dimensional vector of nuclear coordinates.
If the matrix element $X_{JK}$ satisfies translational invariance, 
it follows that:
\begin{align}
\label{xtrans}
 X_{JK}(\vec{R} + \vec{\Delta}) =  X_{JK}(\vec{R})
\end{align}
A first order expansion then yields: 
\begin{align}
\sum_{A,\beta} \frac{\partial X_{JK}}{\partial R_{A \beta}} \Delta_{\beta}= 0 
\end{align}
Thus, if we choose, e.g., ${\bf \Delta}$ to have only a component in only the $\alpha$ direction,
$\Delta_{\beta} = \delta_{\alpha \beta}$,  it follows that our derivative couplings would satisfy
\begin{align}
\label{transzero}
\sum_{A} \frac{\partial X_{JK}}{\partial R_{A \alpha}} = 0 
\end{align}
\noindent and Eq. \ref{zero}. Surface hopping should conserve linear momentum.


A similar thought experiment demonstrates that  conservation of angular momentum should arise (in principle) from the isotropy of space and the derivative coupling should satisfy:
\begin{align}
\sum_{A,\alpha, \beta} 
\epsilon_{\alpha \gamma \beta}
d_{JK}^{A \alpha}   R_{A \beta} =0 
\label{shouldrot}
\end{align}
To that end, suppose that we have a rotationally invariant set of basis functions (and later we will show how construct such a set).  Let $\hat{U}$ be the $3 \times 3$ rotation matrix that rotates the molecular system. 
If the system is indeed rotationally invariant, then we must have: 

\begin{align}
\label{xrot}
 X_{JK}(\hat{U} {\bf R}_1, \hat{U} {\bf R}_2,\ldots, \hat{U} {\bf R}_N) =    X_{JK}( {\bf R}_1,  {\bf R}_2,\ldots,  {\bf R}_N)  
\end{align}
Now, mathematically, such rotational invariance can  be defined only locally--because there are many (different) paths by which one can rotate one direction into another direction in three dimensions. To that extent, it will be very helpful to expand the rotation matrix $\hat{U}$ in terms of the angular momentum operators $(\hat{L}_x, \hat{L}_y, \hat{L}_z)$ around the identity:
\begin{align}
\hat{U} &= \sum_{\gamma}\exp\left(i {\hat{L}^\gamma \delta / \hbar}\right) \\
 &\approx \hat{I} + \frac{i}{\hbar}\sum_{\gamma}\hat{L}^\gamma \delta 
\end{align}
Then, by a first order analysis, using the matrix elements of angular momentum in three dimensions, $\left( \hat{L}^{\gamma} \right)_{\alpha \beta} = i\hbar \epsilon_{\gamma \alpha \beta}$, it follows that:

\begin{align}
\frac{i}{\hbar} \delta \sum_{A,\alpha} \frac{\partial X_{JK}}{\partial R_{A \alpha}} \left( \hat{L}^{\gamma} \Rb_A  \right)_{\alpha} & = 0 
\end{align}
\noindent i.e.
\begin{align}
\sum_{A,\alpha, \beta} \frac{\partial X_{JK}}{\partial R_{A \alpha}} \epsilon_{\alpha \gamma \beta}  R_{A \beta} =0 
\label{rotzero}
\end{align}
In other words, surface hopping should conserve angular momentum.

Interestingly, note that, in the discussion above, one has the freedom to choose any origin. Mathematically, if the matrix elements $X_{JK}$ satisfy Eq. \ref{transzero} as well as Eq. \ref{rotzero}, they will also satisfy
\begin{align}
\label{no_origin}
 \sum_{A,\alpha, \beta} \frac{\partial X_{JK}}{\partial R_{A \alpha}} \epsilon_{\alpha \gamma \beta} \left( R_{A \beta}  -  Q_{\beta} \right) = 0
\end{align}
In other words, one can shift the origin by any arbitrary vector ${\bf Q}$ and still conserve angular and linear momentum.

\subsection{Explicit Form and Properties of the Derivative Coupling}
From the theory presented above, it follows that surface hopping would conserve the nuclear linear and angular momentum if $(i)$ the derivative coupling were of the form of a gradient of a quantify $X$ as in Eq. \ref{xderv}; if $(ii)$ the quantity $X$ were to satisfy translational invariance as in Eq. \ref{xtrans}; if $(iii)$ the quantity $X$ were to satisfy rotational invariance as in Eq. \ref{xrot}.
The fact that surface hopping does not guarantee either nuclear linear or angular momentum conservation implies that at least one of the three conditions above are not met.

At this point, in order to understand how and why surface hopping does not guarantee either nuclear linear or angular momentum conservation, it will be helpful to review the final expression for the derivative coupling that one recovers when applying Hellman-Feynman theory (Eq. \ref{hellman}) in the context of modern electronic structure theory using an atomic orbital basis. In such a case, the form of a derivative coupling is as follows:

\begin{align}
\label{longd}
    d_{JK}^{B\alpha} = \frac{1}{E_K - E_J} \left(\sum_{\mu \nu} \Gamma^{(h)JK}_{\nu \mu} \frac{\partial h_{\mu \nu}}{\partial R_{B \alpha}}  + \sum_{\mu \nu \lambda \sigma} \Gamma^{(\pi)JK}_{\nu \mu \lambda \sigma} \frac{\partial \pi_{\mu \nu  \sigma \lambda }}{\partial R_{B \alpha}}  + \sum_{\mu \nu} \Gamma^{(S)JK}_{\nu \mu} \frac{ \partial S_{\mu \nu}}{\partial R_{B \alpha}}\right) + \sum_{\mu \nu} D^{JK}_{\nu \mu} \tilde{S}_{\nu \mu}^{B \alpha}
\end{align}
Here, $h_{\mu \nu}$ and    $\pi_{\mu \nu \lambda \sigma}$ are (respectively) the one and two electronic Hamiltonian matrices expressed in an atomic orbital basis.  $S_{\mu \nu}$ is the overlap matrix, $S_{\mu \nu} = \left< \mu \middle | \nu \right>$.
The anti-symmetric derivative of the overlap matrix also appears:
\begin{align}
\label{tildeS}
  \tilde{S}^{B\alpha}_{\mu \nu} = \frac{1}{2} \left(
\left< \mu \middle | \frac{\partial \nu  }{\partial R_{B \alpha}} \right>- 
\left<  \frac{\partial \mu  }{\partial R_{B \alpha}} \middle| \nu
\right>  
  \right)
\end{align}

The exact expressions for the reduced matrices $\Gamma^{(h)JK}_{\mu \nu}, \Gamma^{(\pi)JK}_{\mu \nu \lambda \sigma},$ and $\Gamma^{(S)JK}_{\mu \nu}$ depend on the electronic structure method.
In the simplest case, suppose we have a CIS or TDDFT/TDA wavefunction, 
\begin{align}
  \ket{\Psi_J} = \sum_{ia} t^{Ja}_i \ket{\Phi_i^a}
\end{align}
each Slater determinant is composed of molecular orbitals,
\begin{align}
    \ket{\Phi_i^a} = a_a^{\dagger}a_i \ket{\phi_1 \phi_2 \ldots \phi_N}
\end{align}
and each molecular orbital is in turn a combination of atomic orbitals
\begin{align}
\label{molorb}
    \ket{\phi_j}  = \sum_{\mu} C_{\mu j} \ket{\chi_{\mu}}.
\end{align}
In such a case, expressions for the density matrices can be found in Refs. \citenum{fatehi:2011:dercouple,ou:2014:tddft_tda}. Expressions for the 
 derivative couplings in the context of multireference methods can be found in Refs. \citenum{yarkony:1984:jcp_dercouple,yarkony:1984:emailme}, Expressions hold for the TD-DFT/RPA formalisms exist as well
\cite{herbert:2014:jcp_dercouple,liu:2014:dercouple_div,ou:2015:dercouple_div} if one makes the pseudo-wavefunction approximation.

Quite generally,  for all electronic structure approaches, the form of
$D_{\mu \nu}^{JK}$ is quite simple. $D_{\mu \nu}^{JK}$ is the one-electron reduced transition density matrix between states $J$ and $K$:

\begin{eqnarray}
    D_{\mu \nu}^{JK} = \sum_{pq} C_{\mu p} C_{\nu q} \left< \Psi_J \middle| a_p^{\dagger} 
     a_q \middle|
     \Psi_K
    \right>
\end{eqnarray}

Let us now address the three conditions we discussed above as needed for nuclear linear and angular momentum conservation:
\begin{itemize}
\item $(i)$ From Eq. \ref{longd}, it becomes clear that the derivative coupling is not a derivative of the form of Eq. \ref{xderv} or a sum of such derivatives with factors. The problem at bottom is the last term, $\sum_{\mu \nu} D^{JK}_{\nu \mu} \tilde{S}_{\nu \mu}^{B \alpha}$, because that term is clearly not a derivative, as shown in Eq. \ref{tildeS}

\item $(ii)$ All of the other matrix elements in Eq. \ref{longd} (specifically, 
$\frac{\partial h_{\mu \nu}}{\partial R_{B \alpha}}$, $\frac{\partial \pi_{\mu \nu  \sigma \lambda }}{\partial R_{B \alpha}},$ and $\frac{ \partial S_{\mu \nu}}{\partial R_{B \alpha}}$)  do satisfy translation invariance. For example,
\begin{align}
\label{hzero}
\sum_{A} \frac{\partial h_{\mu \nu}}{\partial R_{A \alpha}} = 0 
\end{align}
Thus, it would appear that enforcing linear momentum conservation should be somewhat approachable within the context of a surface hopping calculation.

\item $(iii)$ None of the matrix elements in Eq. \ref{longd} satisfy rotational invariance. For example, 
\begin{align}
\sum_{A,\alpha, \beta} \frac{\partial h_{\mu \nu}}{\partial R_{A \alpha}} \epsilon_{\alpha \gamma \beta} R_{A \beta}  \ne 0
\end{align}
Thus, addressing angular momentum within a FSSH calculation looks to be far more involved than the case of linear momentum.
\end{itemize}

From the considerations above, it will make sense to address linear and angular momentum conservation separately.

\section{Restoring Conservation of Linear Momentum By Accounting For Electronic Linear Momentum}\label{sec-lin}

Formally, the Born-Oppeneheimer treatment is not well-defined without phase conventions for the electronic states at every nuclear geometry, $\vec{R}$.  Quite generally, the standard choice of phase conventions is to choose the electronic states to satisfy:
\begin{align}
\label{pnpezero}
\frac{\hbar}{i}\left(
 \sum_i \frac{\partial}{\partial \bfr_i} +
  \sum_n \frac{\partial}{\partial \Rb_n}
  \right) \Phi_K(\vec{r}; \vec{R})= 
\left(\hat{{\bf P}}_N + \hat{{\bf p}}_e \right) \Phi_K(\vec{r}; \vec{R}) = 0
\end{align}
In words, according to Eq. \ref{pnpezero}, the adiabatic electronic states are identical if one translates both the electronic and nuclear degrees of freedom at the same time, which is equivalent to solving the electronic Schrodinger equation in a so-called space fixed frame (SFF)\cite{yarkony:1989:jcp_emailme_ang}. 
According to this phase convention,  in a basis of Born-Oppenheimer electronic states $\left\{ \ket{\Phi_J(\vec{R})} \right\}$ , the total (nuclear + electronic) linear momentum operator is represented by just the nuclear momentum operator:
\begin{align}
\left(\hat{{\bf P}}_N + \hat{{\bf p}}_e \right) \chi_J(\vec{R}) \Phi_K(\vec{r}; \vec{R}) = 
\left(\hat{{\bf P}}_N \chi_J(\vec{R})\right) \Phi_K(\vec{r}; \vec{R})
\label{littlep}
\end{align}
This relationship is very convenient as far as understanding a trajectory moving along a single surface.

Notwithstanding the beauty of Eq. \ref{littlep}, the phase convention  in Eq. \ref{pnpezero} makes clear that the derivative couplings cannot satisfy Eq. \ref{zero}. In fact,  if we multiply Eq. \ref{pnpezero} by  $\Phi_J\left(\vec{r}, \vec{R}\right)$ and integrate over $\vec{r}$, we find:
\begin{align}
    \sum_B d_{JK}^{B \alpha} = -\sum_i \left< \Phi_J \middle| \frac{\partial }{\partial r_{i\alpha}} \Phi_K \right>
\end{align}
which is not compatible with Eq. \ref{zero}.
More generally,  the implications of Eq. \ref{pnpezero} must be taken seriously in the context of a surface hopping calculation. Consider again the hypothetical H-atom $3p_x\rightarrow 1s$ hop described above for which surface hopping fails.
 Physically, Eq. \ref{pnpezero} stipulates the concurrent motion of electrons and their corresponding nuclei, and Eq. \ref{littlep} dictates that the nuclear momentum operator actually represents the total momentum within a Born-Oppenheimer calculation.   However,  Eqs. \ref{hop2} and \ref{rescale} do not take into account this concurrence; these equations ignore the fact that an electron in a $3p_x$ orbital is dragged along with the proton so that both have the same initial velocity, and the total center of mass for the H-atom (electronic plus nuclear) should function as a constant of motion.
As a result, the hopping rate has the spurious feature whereby the overall translational motion of a collection of nuclei can induce an electronic transition, even when the electrons have the same initial and final momentum.

This line of thinking makes one realize that electronic motion  must be taken into account when calculating a hopping rate.  The usual approach towards solving these problems is to introduce electron translation factors\cite{bates:1958:etf,schneiderman:1969:pr:etf,delos:1978:pra:etf,delos:1981:rmp,winter:1982:pra:etf,errea:1994:etf,ohrn:1994:rmp:etf,riera:1998:PRL,fatehi:2011:dercouple}, which we will now review. An alternative (but perturbatively equivalent) approach based on a different partitioning of the Born-Oppenheimer Hamiltonian (that eventually leads to phase space surface hopping) is described in the Appendix.


\subsection{Electron Translation Factors}

Consider a single atomic  orbital $\chi_{\mu}(\bfr)$ that is  attached to atom $B$ that we intend to displace. The idea of electron translation factors is that we replace:
\begin{eqnarray}
\label{replace}
    \chi_{\mu}(\bfr) \rightarrow \chi_{\mu}(\bfr) \exp\left(\frac{im_e \bfv_B \cdot \bfr} {\hbar}\right)
\end{eqnarray}
This replacement establishes a dynamical phase factor for each electronic orbital and captures the idea that an electron is inevitably pulled along with the nucleus  to which it is centered. This dragging of the orbital leads to a change in the hopping rate because $H_{JK}$ in Eq. \ref{hop} is no longer diagonal; an off-diagonal term has appeared.   

To quantify this effect, note that we will need to make several approximations. 
\begin{enumerate}
\item First, note that if we make the replacement in Eq. \ref{replace}, then the molecular orbitals in Eq. \ref{molorb} need not be orthonormal any longer. We will ignore this change. More generally, we will ignore the difference in phases applied to different atoms. 

\item Second, we will eventually take the limit that $\exp\left(\frac{im_e \bfv_B \cdot \bfr}{\hbar}\right)\rightarrow 1 $. As such, unless we pull down a factor of $\bfv_B$, we will assume all other matrix elements are unchanged (i.e. $\hat{H}$ will remain diagonal).  The primary off-diagonal term then arises from the operation of the electronic kinetic energy operator ($\frac{1}{2m_e}\hat{\bf p} \cdot \hat{\bf p} = \frac{-\hbar^2}{2m_e}\hat{\mathbf{\nabla}}^2$ in Eq. \ref{hel}). 
When we operate $\hat{\mathbf{\nabla}}^2$ on the atomic orbital $\chi_\mu(\bfr)$ (which is associated with nucleus $B$), we find (to first order):
\begin{align}
\label{der2a}
\nabla^2 \chi_\mu = \frac{2im_{e}}{\hbar} {\bf \nabla} \chi_\mu \cdot {\bf v}_B,
\end{align}

i.e.
\begin{align}
\label{der2}
\frac{-\hbar^2}{2m_e} \nabla^2 \chi_\mu = - i \hbar {\bf \nabla} \chi_\mu \cdot {\bf v}_B,
\end{align}

\item Having found that the substitution in Eq. \ref{replace} leads to a new hopping rate (and that we must invoke Eq. \ref{hop} instead of Eq. \ref{hop2}), it is fairly straightforward to estimate the total new matrix element between adiabats $J$ and $K$ because the electronic kinetic energy operator is a one-electron operator.  In such a case,  we require only the matrix elements between single electron states. Following Eq. 
\ref{der2}, the matrix elements are of form:
\begin{align}
\tilde{\zeta}_{\nu \mu} &= -\int d{\bf r} \chi_{\nu} {i \hbar} \left( {\bf \nabla} \chi_\mu \cdot {\bf v}_B\right) \\
& = -\sum_{\alpha}  i \hbar v_{B \alpha} \int d{\bf r} \; \chi_{\nu} \; \frac{\partial \chi_\mu}{\partial r_{\alpha}}
\end{align}
At this point, we note that, by translational invariance and the proper choice of electronic phase, one expects that moving the atomic orbitals (electrons and nuclei) should not introduce a new phase to the wavefunction. Formally, consider an orbital $\chi_{\mu}$ defined centered at a fixed nuclear position $\Rb^0_B$. If we we wish to transport this same orbital so that the orbital is now centered at a nearby nuclear position $\Rb^B$, the relevant formula is:
\begin{align}
\label{exp-pbasis}
    \ket{\chi_\mu(\Rb_B)} =   e^{(-i\hat{\bf p}_e)\cdot(\Rb_B-\Rb_B^0)/\hbar  }\ket{\chi_\mu(\Rb_B^0)}
\end{align}
Thus, 
\begin{align}
\label{translate_ao0}
    \left( \hat{\Pb}_N + \hat{\bf p}_e \right) \ket{\chi_\mu(\Rb_B)} =   0
\end{align}
or in other words,
\begin{align}
\label{translate_ao}
\left(\frac{\partial}{\partial r_{\alpha}} + \frac{\partial}{\partial R_{B \alpha}}\right) \chi_{\mu}({\bf r};\Rb_B) = 0,
\end{align} 
Eq. \ref{translate_ao} is the single-orbital equivalent of the more formal many-body quantum mechanics in Eq. \ref{pnpezero}.
From Eq. \ref{translate_ao}, it follows that:
\begin{align}
\tilde{\zeta}_{\nu \mu} 
& = \sum_{\alpha}  {i \hbar} v_{B \alpha} \int d{\bf r} \; \chi_{\nu} \; \frac{\partial \chi_\mu}{\partial R_{B \alpha}}
\end{align}
Finally, it is important to recognize that if the atomic orbitals $\chi_{\mu}$ and $\chi_{\nu}$ were orthonormal, then we would have (for all $B$ and $\alpha$):
\begin{align}
\int d{\bf r} \; \left( \chi_{\nu} \; \frac{\partial \chi_\mu}{\partial R_{B \alpha}} + 
 \frac{\partial \chi_\nu}{\partial R_{B \alpha}} \; \chi_{\mu} 
\right) = 0
\end{align}
Thus, the matrix $\tilde{\zeta}$ would be hermitian. That being said, the atomic orbitals are not orthonormal and so, in order for this matrix to represent a Hamiltonian operator, it makes sense to``hermitianize'' the matrix:
\begin{align}
\zeta_{\nu \mu} &= 
\frac{1}{2}\left( \tilde{\zeta}_{\nu \mu} + \tilde{\zeta}_{\mu \nu}^* \right) \\
& = \sum_{\alpha}  \frac{i \hbar}{2} v_{B \alpha} \int d{\bf r} \; \left( \chi_{\nu} \; \frac{\partial \chi_\mu}{\partial R_{B \alpha}}  - \frac{\partial \chi_\nu}{\partial R_{B \alpha}} \; \chi_{\mu}   \right) \\
&= {i \hbar} \sum_{\alpha} v_{B \alpha} \tilde{S}^{B \alpha}_{\nu \mu} 
\end{align}
where we have invoked the anti-symmetric overlap derivative from Eq. \ref{tildeS}.
\end{enumerate}

With the three assumptions above, the final expression for the Hamiltonian matrix element can be written cleanly in terms of the one electron reduced transition density matrix $(D^{JK}_{\mu \nu})$. If we allow all nuclei to move, the final result is:
\begin{align}
\label{newJK}
H_{JK} &= \sum_{\mu \nu}  D^{JK}_{\mu \nu} \zeta_{\nu \mu} 
= i \hbar \sum_{\mu, \nu, B, \alpha} 
D^{JK}_{\mu \nu} v_{B \alpha} \tilde{S}^{B \alpha}_{\nu \mu} 
\end{align}
The final hopping probability (Eq. \ref{hop}) becomes:
\begin{align}
    \Gamma_{J\rightarrow K} &= 2\Re \left( \frac{ \left[ \frac{i}{\hbar}H_{JK} + \sum_{B \alpha} d^{B\alpha}_{JK}v^{B \alpha} \right]\rho_{KJ}}{\rho_{JJ}}\right) 
    \label{dressedGamma}\\
    & = 2\Re \left( \frac{  \sum_{B \alpha} \left( d^{B\alpha}_{JK} - \sum_{\mu \nu}D^{JK}_{\mu \nu} \tilde{S}^{B \alpha}_{\nu \mu} \right)v^{B \alpha} \rho_{KJ}}{\rho_{JJ}}\right)
        \label{dressedGamm2}
\end{align}

Let us now revisit the case of hydrogen atom, where the full electronic state is nothing more than an atomic orbital. We can now prove that the analysis above completely eliminates all hopping between energy levels. To see this point, let us choose $J = \chi_{\mu} = \phi_g({\bf r})$ to be the $1s$ electronic ground state; let $K = \chi_{\nu} = \phi_e({\bf r})$ be an excited $p$ electronic orbital. If we are interested in the transition between these two states, we can set $D^{JK}_{\mu \nu} = 1$. Now, the two states of interest are orthogonal ($\left<\phi_g \middle| \phi_e \right> = 0$) which implies that 
$\left<\phi_g \middle|  \frac{\partial}{\partial  R_{\alpha}} \phi_e \right> + \left< \frac{\partial}{\partial  R_{\alpha}}\phi_g \middle|   \phi_e \right> = 0$. Hence, it follows that

\begin{align}
d_{ge}^{\alpha} \equiv \left<\phi_g \middle|  \frac{\partial}{\partial  R_{\alpha}} \phi_e \right> = \frac{1}{2} \left(
\left<\phi_g \middle|  \frac{\partial}{\partial  R_{\alpha}} \phi_e \right> - \left< \frac{\partial}{\partial  R_{\alpha}}\phi_g \middle|   \phi_e \right>
\right) = \tilde{S}^{\alpha}_{ge}
\end{align}
and thus (by Eq. \ref{dressedGamm2}) that $\Gamma_{J \rightarrow K} = 0.$

More generally, this exercise introduces the concept of an ETF-corrected derivative coupling that more properly recovers the correct rates of electronic hopping and the correct direction for momentum rescaling:
\begin{align}
\label{detf}
d_{JK}^{ETF, B \alpha} &= d_{JK}^{B\alpha} - \sum_{\mu \nu}D^{JK}_{\mu \nu} \tilde{S}^{B \alpha}_{\nu \mu} 
\end{align}

Note that, according to Eq. \ref{longd}, this ETF-corrected derivative coupling is equivalent to ignoring all terms proportional to $\tilde{S}$:
\begin{align}
\label{longdetf}
    d_{JK}^{ETF,B\alpha} = \frac{1}{E_K-E_J}\left( \sum_{\mu \nu} \Gamma^{(h)JK}_{\nu \mu} \frac{\partial h_{\mu \nu}}{\partial R_{B \alpha}}  + \sum_{\mu \nu} \Gamma^{(\pi)JK}_{\nu \mu \lambda \sigma} \frac{\partial \pi_{\mu \nu  \sigma \lambda }}{\partial R_{B \alpha}}  + \sum_{\mu \nu} \Gamma^{(S)JK}_{\nu \mu} \frac{ \partial S_{\mu \nu}}{ \partial R_{B \alpha}} \right)
\end{align}
Furthermore, as follows from Eq. \ref{hzero} (and the analogous expressions for  $\pi$ and $S$), this direction does not yield any change in to the overall total momentum:
\begin{align}
\label{sumzero}
    \sum_B d_{JK}^{ETF,B\alpha} = \frac{1}{E_K-E_J}\left( \sum_{\mu \nu B} \Gamma^{(h)JK}_{\nu \mu} \frac{\partial h_{\mu \nu}}{\partial R_{B \alpha}}  + \sum_{\mu \nu B} \Gamma^{(\pi)JK}_{\nu \mu \lambda \sigma} \frac{\partial \pi_{\mu \nu  \sigma \lambda }}{\partial R_{B \alpha}}  + \sum_{\mu \nu B} \Gamma^{(S)JK}_{\nu \mu} \frac{ \partial S_{\mu \nu}}{ \partial R_{B \alpha}}\right) =0
\end{align}
Eq. \ref{sumzero} can also be derived directly from Eq. \ref{pnpezero}, using basic electronic structure theory manipulations:
\begin{align}
\label{sumzero1}
    \sum_B d_{JK}^{B \alpha} & = -\sum_i \left< \Phi_J \middle| \frac{\partial }{\partial r_{i \alpha}} \Phi_K \right> \\
    \label{sumzero2}
    & =-\frac{1}{2}  \sum_i \left( \left< \Phi_J \middle| \frac{\partial }{\partial r_{i \alpha}} \Phi_K \right>  - \left< \frac{\partial }{\partial r_{i \alpha}} \Phi_J \middle|  \Phi_K \right>\right) \\
    \label{sumzero3}
      & =-\frac{1}{2}  \sum_{\mu \nu} D_{\mu \nu}^{JK} \left( \left< \chi_\mu \middle| \frac{\partial }{\partial r_{\alpha}} \chi_\nu \right>  - \left< \frac{\partial }{\partial r_{\alpha}} \chi_\mu \middle|  \chi_\nu  \right>\right) \\  
      \label{sumzero4}
          & =\frac{1}{2}  \sum_{\mu \nu B} D_{\mu \nu}^{JK} \left( \left< \chi_\mu \middle| \frac{\partial }{\partial R_{B\alpha}} \chi_\nu \right>  - \left< \frac{\partial }{\partial R_{B\alpha}} \chi_\mu \middle|  \chi_\nu  \right>\right) \\
          \label{sumzero5}
          &=\sum_{\mu \nu B}D^{JK}_{\mu \nu} \tilde{S}^{B \alpha}_{\nu \mu} 
\end{align}
Here, between Eqs. \ref{sumzero1} and \ref{sumzero2}, we have used the fact that $\frac{\partial}{\partial r_{i \alpha}}$  is an anti-Hermitian operator. In Eq. \ref{sumzero3}, we have converted the calculation to a second-quantized form using the fact that $\frac{\partial}{\partial r_{\alpha}}$ is a one-electron operator. In Eq. \ref{sumzero4}, we have used Eq. \ref{translate_ao}. Eq. \ref{sumzero5} follows from the definition in Eq. \ref{tildeS}.

It is important to emphasize that Eq. \ref{detf} is not equivalent to simply removing the translational component of the derivative coupling:
\begin{align}
\label{dnot}
d_{JK}^{ETF, B \alpha} \ne d_{JK}^{B\alpha} -
\frac{1}{N} \sum_{B} d_{JK}^{B\alpha}
\end{align}
Defining ETFs as in Eq. \ref{dnot} would not be very physical because the expression is not size consistent.  More precisely, consider two non-interacting systems 1 and 2, and an excitation between states $J$ and $K$ in system 1, so that $d_{JK}^{B \alpha}$ is nonzero exclusively for atoms $B$ on system 1. According to the definition Eq. \ref{dnot}, $d_{JK}^{ETF, B \alpha}$ would be nonzero for atoms $B$ in either system 1 or 2 -- which does not make sense.
At a minimum, one would expect that a more physical, size-consistent correction would be:
\begin{align}
\label{dnot2}
d_{JK}^{ETF, B \alpha} \stackrel{?}{=} d_{JK}^{B\alpha} -
\frac{\left|d_{JK}^{B\alpha}\right|}{\sum_{A} \left|d_{JK}^{A\alpha}\right|}
\sum_{A} d_{JK}^{A\alpha}
\end{align}
Note that this expression still does not agree with Eq. \ref{detf}.
See Appendix \ref{tian}.

\section{Restoring Conservation of Angular Momentum By Accounting For Electron Angular  Momentum} \label{sec-ang}

All of the theory presented above for the case of linear momentum should carry over to the case of angular momentum.  After all, both linear and angular momenta are equally good quantum numbers. And yet, there are two significant differences between these two quantities. First, it is well known that one cannot fully separate angular momentum from vibrations for non-rigid bodies; Coriolis forces inevitably arise. Thus, if one were able to remove all angular momentum components of the derivative coupling, in the process one would also necessarily change (though likely not too much) some vibrational components of the derivative coupling.

Second, all electronic structure calculations are performed in a space fixed frame where the basis functions are assumed to translate with nuclear motion (and not rotate). In other words, even though the many body wavefunctions must follow \cite{littlejohn:2023:ang},

\begin{align}
\left(\hat{{\bf L}}_N + \hat{{\bf L}}_e \right) \Phi_K(\vec{r}; \vec{R}) = 0,
\end{align}
a single-particle atomic orbital basis (parameterized by a single atom $B$) does not satisfy any analogue of Eq. \ref{translate_ao}:
\begin{align}
\label{rotate_ao}
\sum_{\alpha \beta \gamma} \epsilon_{\alpha \beta \gamma} \left( r_{\beta}\frac{\partial}{\partial r_{\alpha}} + R_{B\beta}\frac{\partial}{\partial R_{B \alpha}}\right) \chi_{\mu}(\vec{r};\vec{R}) \ne 0,
\end{align} 
Thus, if one wished to include ``electronic rotational factors'' or ERFs so as to allow for electronic orbitals to rotate (rather than just translate) with their associated nucleus, the task would not be very easy.

Now, given these complications, the most obvious fix for enforcing the conservation of angular momentum is simply to remove the offending component of the angular momentum. {One approach is the  projection scheme used in Ref.~\citenum{truhlar:2020:project_rot_nac}.  This scheme can be derived most easily by fixing the origin to be the ``centroid'' (i.e. $\sum_AR_A=0$).  We derive this result in a very general form in Appendix~\ref{tian} using the method of Lagrange multipliers and show that, after removing  the translational and rotational components of the derivative coupling, the final result is:
\begin{align}
    \label{eq-truhlar}
    d^{A\alpha}_{ETF+ERF} = d^{A\alpha} - \frac{1}{N}\sum_{B}d^{B\alpha}-\sum_{\beta\gamma}\epsilon_{\alpha\beta\gamma}R_{A\alpha}\sum_{\beta'}\mathcal{M}^{-1}_{\beta\beta'}\sum_B\sum_{\alpha'\gamma'}\epsilon_{\alpha'\beta'\gamma'}R_{B\gamma'}d^{B\alpha'},
\end{align}
where $N$ is the total number of nuclei, and $\mathcal{M}$ is the moment of inertia matrix with all the masses set to 1.}
However, this equation is difficult to justify.  In particular, just like Eq. \ref{dnot}, this approach yields non-size consistent derivative couplings. As discussed above, consider two infinitely separated systems, 1 and 2, and two states ($J$ and $K$) localized to system 1.  The derivative coupling $d_{JK}^{B \alpha}$ will also be localized to system 1. Unfortunately, however, Eq. \ref{eq-truhlar} will yield a dressed derivative coupling $d_{JK}^{ETF-rot, B \alpha}$ delocalized over both systems 1 and 2. For this reason, there is good reason to derive a correction beyond Eq. \ref{eq-truhlar} for removing the angular momentum component from the derivative coupling in a size-consistent, physical fashion.

\subsection{A rotationally boosted set of atomic orbitals}

Guided by our analysis above that dictated we work in a translationally boosted basis, we will presently aim to work a basis that is rotating which should enable us to remove the rotationally problematic components of the derivative coupling. Now, if all  basis functions were s-orbitals, then our basis would automatically be rotationally invariant: after all, an $s$ orbital is an $s$ orbital in any frame.  The problem is, however, that all practical quantum chemistry calculations involve (at least) $p$ and $d$ orbitals on atomic sites and these orbitals do not individually transform in a  rotationally invariant fashion.  For the most part, one never cares about the orientation of these basis functions. After all, the electronic energy from the Schrodinger equation is invariant to the identity of the individual basis functions; the energy depends only on the overall vector space of basis functions. That being said, the expression in Eq. \ref{longdetf} does care about the definition of each atomic orbital $\chi_{\mu}$ and so we must be careful as we seek to construct rotationally boosted atomic orbitals $\tilde{\chi}_{\mu}$.





To begin our analysis, suppose we have an arbitrary nucleus at ${\bf R}_0$ and consider a rotation  to
\begin{align}
{\bf R} = 
\hat{U} {\bf R}_0 = e^{\frac{i}{\hbar}\sum_{\gamma}\hat{L}_n^{\gamma} \theta_{\gamma}
} {\bf R}_0.
\end{align}
For now, we will not specify the origin of the rotation; as discussed around Eq. \ref{no_origin}, if one seeks to find derivative couplings that satisfy Eqs. \ref{zero} and \ref{shouldrot}, one has freedom in choosing the origin. 
We can construct a proper rotationally invariant basis $\ket{\tilde{\chi}_{\mu}}$ at position ${\bf R}$ from the basis functions at $\ket{\chi_{\mu}}$ at position ${\bf R}_0$ as follows:

\begin{align}
\label{explbasis}
    \ket{\tilde{\chi}_{\mu} ({\bf R})} = e^{-\frac{i}{\hbar}\sum_{\gamma}  \hat{L}_e^{\gamma} \theta_{\gamma}} \ket{\chi_{\mu} ({\bf R}_0)}
\end{align}


Next, we expand the rotation $\hat{U}$ as an infinitesimal displacement from the identity: 
\begin{align}
\label{little}
{\bf R}_B = \left(I - \frac{i}{\hbar}\sum_{\alpha} \theta_{\alpha} \hat{L}^{\alpha}\right) {\bf R}_B^0 
\end{align}

Formally, the equation above is  an overdetermined set of equations for $\theta(\vec{R})$ because one does not constrain onself to move along a rotational degree of freedom only. Nevertheless, we can recover a least squares fit for the solution by introducing weights and following standard linear algebra inversion routines.
Let us define $\mathcal{R}$ be the matrix of positions of each of the nuclei weighted by some atom-specific parameter $\zeta$, 
\begin{align}
\label{sqrtm}
    \mathcal{R} = \left[ \; \sqrt{\zeta_1} \Rb_1,  \sqrt{\zeta_2} \Rb_2, \ldots , \; \sqrt{\zeta_N} \Rb_N \right].
\end{align}  
According to Eq. \ref{little}, 
\begin{align}
& {\mR - \mR^0} =  -\frac{i}{\hbar}\sum_{\alpha} \theta_{\alpha} \hat{L}_{\alpha} \mR^0 \\
\implies
& \left(\mR - \mR^0\right) (\mR^0)^T \left(\mR^0(\mR^0)^T\right)^{-1} =  -\frac{i}{\hbar}\sum_{\alpha} \theta_{\alpha} \hat{L}^{\alpha} 
\end{align}
In other words, 
\begin{align}
\mR  (\mR^0)^T \left(\mR^0(\mR^0)^T\right)^{-1} -I =  -\frac{i}{\hbar}\sum_{\alpha} \theta_{\alpha} \hat{L}^{\alpha} 
\end{align}
Here, we see the introduction of the  tensor:
\begin{align}
\Lambda =  \mR^0(\mR^0)^T
\end{align}
or in coordinates, 
\begin{align}
\Lambda_{\alpha \beta} =  \sum_B \zeta_B \Rb^0_{B \alpha}
\Rb^0_{B \beta}
\end{align}
Finally, using $\tr(\hat{L}^{\alpha}) = 0$
and $\tr(\hat{L}^{\alpha} \hat{L}^{\beta}) = 2i \hbar^2 \delta_{\alpha \beta}$, it follows that:
\begin{align}
\label{theta1}
\theta_{\alpha} &=
-\frac{1}{2 \hbar} \tr \left(
 \mR  (\mR^0)^T \left(\mR^0(\mR^0)^T \right)^{-1} \hat{L}_{\alpha}
\right) \\
 &= -\sum_{\beta, \gamma, \tau, B}
 \frac{1}{2 \hbar} \zeta_B R_{B \beta} R^0_{B \tau} \Lambda^{-1}_{\tau \gamma} \epsilon_{\alpha \beta \gamma}
 \label{theta2}
\end{align}

Thus, at the end of the day, we can write a rotating basis as
\begin{align}
\label{tilde1}
    \ket{\tilde{\chi}_{\mu} ({\bf R})} &= \exp \left(
  \sum_{\alpha, \beta, \gamma,  \delta, B}
 \frac{-i}{2 \hbar} \zeta_B R_{B \beta} R^0_{B \delta} \Lambda^{-1}_{\delta \gamma} \epsilon_{\alpha \beta \gamma} \hat{L}_e^{\alpha}  
    \right) \ket{\chi_{\mu} ({\bf R}_0)},
\end{align}
or, in terms of the displacement ${\bf R} = {\bf R_0} + {\eta}$:
\begin{align}
\label{tilde1b}
    \ket{\tilde{\chi}_{\mu} ({\bf R})} &= \exp \left(
  \sum_{\alpha, \beta, \gamma, \delta, B}
 \frac{-i}{2 \hbar} \zeta_B \eta_{B \beta} R^0_{B \delta} \Lambda^{-1}_{\delta \gamma} \epsilon_{\alpha \beta \gamma} \hat{L}_e^{\alpha}  
    \right) \ket{\chi_{\mu} ({\bf R}_0)}.
\end{align}

\noindent It follows that to first order in $\eta$: 
 \begin{align}
 \label{tilde2}
\ket{\tilde{\chi}_{\mu} ({\bf R})} & = \ket{\chi_{\mu} ({\bf R}_0)} - \sum_{\alpha, \beta, \gamma, \delta, B}
 \frac{i}{2\hbar} \zeta_B R_{B \beta} R^0_{B \delta} \Lambda^{-1}_{\delta \gamma} \epsilon_{\alpha \beta \gamma} \hat{L}_e^{\alpha} \ket{\chi_{\mu} ({\bf R}_0)} + O(\eta^2)
 \\
  \label{tilde2b}
 & = \ket{\chi_{\mu} ({\bf R}_0)} - \sum_{\alpha, \beta, \gamma, \delta, B}
 \frac{i}{2\hbar} \zeta_B \eta_{B \beta} R^0_{B \delta} \Lambda^{-1}_{\delta \gamma} \epsilon_{\alpha \beta \gamma} \hat{L}_e^{\alpha} \ket{\chi_{\mu} ({\bf R}_0)} + O(\eta^2)
\end{align}
From this, an easy calculation shows that
\begin{align}
\label{lnlezero}
\left(\hat{{\bf L}}_N + \hat{{\bf L}}_e \right) \tilde{\chi}_\mu({\bf r}; \Rb) = 0 + O(\eta)
\end{align}
and in particular
\begin{align}
\label{lnlezero2}
\left(\hat{{\bf L}}_N + \hat{{\bf L}}_e \right) \tilde{\chi}_\mu({\bf r}; \Rb)\biggr |_{{\bf R} = {\bf R}_0} = 0
\end{align}
More generally, according to Ref. \citenum{littlejohn:2023:ang}, Eq. \ref{lnlezero} should hold rigorously to all orders when evaluated 
at any point configuration $\vec{R}$ that arises from a rigid rotation of the original configuration $\vec{R}_0$ when the correct rotation angle is chosen.




\subsection{The derivative coupling in a rotating frame  }
At this point, we can construct the derivative coupling in a rotated frame using Eq. \ref{tilde2} and take the derivative of the relevant matrix elements. For instance, consider the one-electron Hamiltonian operator $\hat{h}$: 
\begin{align}
    h_{\tilde{\mu} \tilde{\nu}} \left( \vec{R} \right) =
    \left< \tilde{\chi}_{\mu}(\vec{R}) \middle| 
    \hat{h}(\vec{R})
    \middle|
    \tilde{\chi}_{\nu}(\vec{R}) \right>
\end{align}

Using the chain rule, we can identify the key (new) term that arises when we use rotationally boosted basis functions and evaluate the entire expression at $\vec{R} = \vec{R}_0$:
\begin{align}
\frac{\partial h_{\tilde{\mu} \tilde{\nu}}}{\partial R_{A \beta}} 
&=  \frac{\partial h^{op}_{\mu \nu}}{\partial R_{A \beta}}
+  \frac{i}{2 \hbar}\sum_{\alpha \delta \gamma \kappa \tau} \epsilon_{\alpha \beta \gamma}
    \zeta_A R^0_{A \delta} \Lambda^{-1}_{\delta \gamma} 
\left< \chi_{\mu} \middle| \left[ \hat{L}_e^{\alpha}, h \right] \middle|  \chi_{\nu}  \right>
\label{commtermhere}
 \end{align}

Here, we have defined:
\begin{align}
\frac{\partial h^{op}_{\mu \nu}}{\partial R_{A \beta}} = 
\left< \chi_{\mu} \middle| \frac{\partial \hat{h}}{\partial  R_{A \beta} } \middle|  \chi_{\nu}  \right>
 \end{align}

For the second term on the right hand side of Eq. \ref{commtermhere}, note that by the isotropy of space,
\begin{align}
\left[
\hat{h},\hat{L}_e^{\alpha} + \hat{L}_N^{\alpha}
\right] = 0
 \end{align}

 Therefore, the derivative can also be written as:

 \begin{align}
\frac{\partial h_{\tilde{\mu} \tilde{\nu}}}{\partial R_{A \beta}} \biggr |_{{\bf R} = {\bf R}_0} 
& = \frac{\partial h^{op}_{\mu \nu}}{\partial R_{A \beta}}
\biggr |_{{\bf R} = {\bf R}_0}
-  \frac{i}{2 \hbar}\sum_{\alpha \delta \gamma} \epsilon_{\alpha \beta \gamma}
    \zeta_A R^0_{A \delta} \Lambda^{-1}_{\delta \gamma} 
\left< \chi_{\mu} \middle| \left[ \hat{L}_N^{\alpha}, h \right] \middle|  \chi_{\nu}  \right> 
\biggr |_{{\bf R} = {\bf R}_0}
\\ 
     &=  \frac{\partial h^{op}_{\mu \nu}}{\partial R_{A \beta}}
     \biggr |_{{\bf R} = {\bf R}_0}
-  \frac{1}{2}\sum_{\alpha \delta \gamma \kappa \tau} \epsilon_{\alpha \beta \gamma}
    \zeta_A R^0_{A \delta} \Lambda^{-1}_{\delta \gamma}  \sum_B \epsilon_{\alpha \kappa \tau}  R_{B \kappa} \frac{\partial h^{op}_{\mu \nu}}{\partial R_{B \tau}} 
    \biggr |_{{\bf R} = {\bf R}_0}
     \\
    &= \frac{\partial h^{op}_{\mu \nu}}{\partial R_{A \beta}}
    \biggr |_{{\bf R} = {\bf R}_0}
-   \frac{1}{2} \sum_{\delta \gamma} 
     \zeta_A R^0_{A \delta}\Lambda^{-1}_{\delta \gamma}   \sum_B \left(    R_{B\beta} \frac{\partial h^{op}_{\mu \nu}}{\partial R_{B \gamma}} - 
R_{B \gamma} \frac{\partial h^{op}_{\mu \nu}}{\partial R_{B \beta}}
     \right) 
     \biggr |_{{\bf R} = {\bf R}_0}
\label{finaleq-op-almost}
\end{align}
or if we drop the zero subscript,
\begin{align}
\frac{\partial h_{\tilde{\mu} \tilde{\nu}}}{\partial R_{A \beta}}     &= \frac{\partial h^{op}_{\mu \nu}}{\partial R_{A \beta}}
-   \frac{1}{2} \sum_{\delta \gamma} 
     \zeta_A R_{A \delta}\Lambda^{-1}_{\delta \gamma}   \sum_B \left(    R_{B\beta} \frac{\partial h^{op}_{\mu \nu}}{\partial R_{B \gamma}} - 
R_{B \gamma} \frac{\partial h^{op}_{\mu \nu}}{\partial R_{B \beta}}
     \right) 
\label{finaleq-op}
\end{align}


Now, Eq. \ref{finaleq-op}  represents the derivative of the one electron hamiltonian matrix elements in  a rotating basis functions. If one constructs the cross product with the position operator, it is very straightforward to prove that:
\begin{align}
    \sum_{A \alpha \beta} \epsilon_{\alpha \beta \gamma}
     R  _{A \alpha} \frac{\partial h_{\tilde{\mu} \tilde{\nu}}}{\partial R_{A \beta}} =0
     \label{zerotilderot}
\end{align}

Now, we must emphasize that the choice of basis functions in Eq. \ref{explbasis} (which led to Eq. \ref{finaleq-op}) is appropriate only for motion along rotations but not translations and/or strictly internal motion; as discussed above, Eq. \ref{lnlezero} is exact to all orders only along rigid rotations of the original point ${\bf R}_0$.  \footnote{The fact that Eq. \ref{finaleq-op} is relevant only for rotations   becomes obvious when one considers the fact that $\frac{\partial S^{op}_{\mu \nu}}{\partial R_{A \beta}} = \frac{\partial \pi^{op}_{\mu \nu \lambda \sigma}}{\partial R_{A \beta}} = 0$. Thus, if Eq. \ref{explbasis} were relevant more generally, there would be no Pulay terms or two-electron derivative terms.}
In general, given that molecules can translate, rotate and distort, the most straightforward and robust approach is still to translate the individual basis functions for each atom between different geometries, as in Eq. \ref{exp-pbasis}, and then differentiate, thus recovering $\frac{\partial h_{\mu \nu}}{\partial R_{A \beta}}$.
From these considerations, it is clear that we will need to compromise: it is impossible to construct translationally invariant and rotationally invariant one-electron basis functions at all possible geometries in a smooth fashion where each basis function is parameterized by a single nuclear position. 
From the form of Eq. \ref{finaleq-op}, a clear compromise candidate would then be:
 \begin{align}
\frac{\partial h_{\tilde{\mu} \tilde{\nu}}}{\partial R_{A \beta}} 
    & \rightarrow \frac{\partial h_{\mu \nu}}{\partial R_{A \beta}}
-   \frac{1}{2} \sum_{\alpha \delta \gamma} 
     \zeta_A R_{A \delta}\Lambda^{-1}_{\delta \gamma}   \sum_B \left(    R_{B\beta} \frac{\partial h_{\mu \nu}}{\partial R_{B \gamma}} - 
R_{B \gamma} \frac{\partial h_{\mu \nu}}{\partial R_{B \beta}}
     \right) 
\label{finaleqy}
\end{align}
In Eq. \ref{finaleqy}, if we ignore the second term on the right hand side, our result reduces to the standard result in quantum chemistry.  Moreover, when we include the second term, Eq. \ref{zerotilderot} still holds. 

By linearity, we would then predict the following form for a translationally and rotationally invariant derivative coupling:
\begin{align}
d_{ETF+ERF,JK}^{A \beta}
&= d_{ETF,JK}^{A \beta}
-   \frac{1}{2} \sum_{\delta \gamma} 
      \zeta_A R_{A \delta}\Lambda^{-1}_{\delta \gamma}   \sum_B \left(    R_{B\beta} d^{B \gamma}_{ETF, JK}  - 
R_{B \gamma} d^{B \beta}_{ETF, JK} 
     \right) 
     \label{guess}
\end{align}

\subsection{Size Consistency and Electron Rotational Factors}

It remains only to pick a value for the weights $\zeta$ and, if necessary, in the spirit of Eq. \ref{no_origin}, pick an origin.
Let us begin with the latter question. In order for the final derivative coupling to be translationally invariant, according to Eq. \ref{guess}, one can simply set the origin to satisfy:
\begin{align}
  \sum_B \zeta_B \Rb_{B}
 = 0
\end{align}

Finally, let us address the question of how to choose the weights, $\zeta$.  As described many times above, a momentum rescaling direction can only be physical if that direction is size consistent. However, size consistency with Eq. \ref{guess} is not automatic because of the sum over atoms (B) on the second term on the right hand side; this term arises because the value for $\theta$  in Eq. \ref{theta2} (which specifies the angle of rotation) is not atom specific. Intuitively, a rigid rotation involves collectively moving all of the atoms together, but if we work with a one-electron basis where each basis depends on a single nuclear position, one cannot disentangle rotations from other motion.  Eq. \ref{theta2} then approximates the rotation by analyzing the average rotation motion of many nuclei, even though this approach would seem to clearly break size-consistency.

The way out of this dilemma is to choose the weights $\zeta$ in a way that ensures that different, non-interacting subsystems are never entangled. 
Mathematically, this goal can be achieved by choosing 
\begin{align}
\zeta_A = \norm{{\bf d}_{ETF,JK}^A} = 
\sqrt{\left| d_{ETF,JK}^{Ax}\right|^2+
\left| d_{ETF,JK}^{Ay}\right|^2+
\left| d_{ETF,JK}^{Az}\right|^2}.
\end{align}
In such a case, 
our  final expression for the derivative couplings can be written in a very simple form, whereby one can clearly identify the "electronic rotational factors" (ERFs) for removing any angular component of the derivative coupling:
\begin{align}
\label{easy1final}
d_{ETF+ERF,JK}^{A \beta}
&= d_{ETF,JK}^{A \beta}
-   \frac{\norm{{\bf d}_{ETF,JK}^A}}{2} \sum_{\delta \gamma} 
      R_{A \delta}\Lambda^{-1}_{\delta \gamma}   \sum_B \left(    R_{B\beta} d^{B \gamma}_{ETF, JK}  - 
R_{B \gamma} d^{B \beta}_{ETF, JK} 
     \right) 
\end{align}
where $\Lambda$ is

\begin{align}
\Lambda_{\alpha \beta} =  \sum_B \norm{{\bf d}_{ETF,JK}^B} R_{B \alpha}
R_{B \beta}
\end{align}
and the origin has been chosen such that:
\begin{align}
  \sum_B \norm{{\bf d}_{ETF,JK}^B} \Rb_{B}
 = 0
\end{align}

Eq. \ref{easy1final}  is size consistent because if $\norm{{\bf d}_{ETF,JK}^A} = 0$, one is guaranteed that 
$\norm{{\bf d}_{ETF+ERF,JK}^A} = 0$ as well.
Use of Eq. \ref{easy1final} also
ensures that momentum rescaling will conserve the nuclear linear and angular momentum within the surface hopping algorithm. In other words, Eq. \ref{easy1final}  explicitly satisfies:

\begin{eqnarray}
    \sum_A d_{ETF+ERF,JK}^{A \beta}  &=&  0\\
    \sum_{A \alpha \beta} \epsilon_{\alpha \beta \gamma}
     R _{A \alpha} d_{ETF+ERF,JK}^{A \beta}&=& 0
\end{eqnarray}

Note that, as in the ETF case, Eq. \ref{easy1final} cannot be derived through a naive minimization scheme based on Lagrange multipliers. See Appendix~\ref{tian}.

\section{Results} \label{sec-res}

\begin{figure}[h]
    \centering
    \begin{minipage}{0.45\textwidth}
        \centering
        \includegraphics[width=0.99\textwidth]{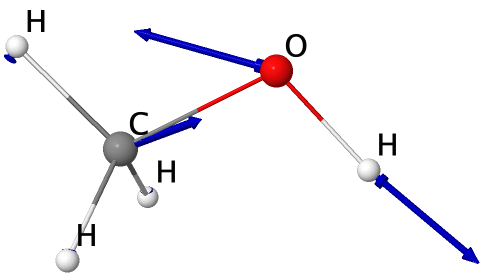}
    \end{minipage}\hfill
    \begin{minipage}{0.45\textwidth}
        \centering
        \includegraphics[width=0.99\textwidth]{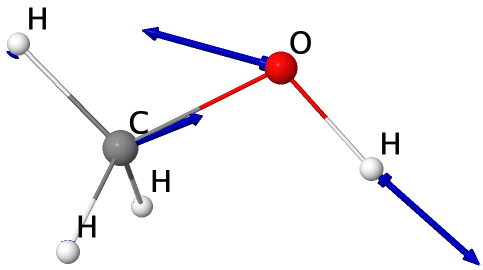}
    \end{minipage}
    
    \caption{Derivative coupling vectors between the 1st and 2nd CIS states for methanol without (left) and with (right) electron rotational factors (ERFs). For these molecular states, there is virtually no difference between the two vectors, as can be confirmed from the values in Table \ref{tab:methanol_dcs}.}
    \label{fig:methanol_dc}
\end{figure}

While ETFs are routinely calculated nowadays by most electronic structure packages that can evaluate derivative couplings, 
Eq. \ref{easy1final} is novel.  One might label these terms as electron rotational factors (ERF's) and to our knowledge, they are not routinely calculated by any electronic structure package.
Using a developmental version of Q-Chem\cite{qchem6}, we have calculated the derivative  couplings between the first and the fourth configuration-interaction singles (CIS) states for the methanol molecule at the CIS/def2-svp level of theory including (or not including) ETFs and/or ERFs. 
The first and the fourth CIS states have excitation energies \qty{8.74}{\electronvolt} and \qty{12.20}{\electronvolt} respectively (see Appendix for the geometry used).
In Table \ref{tab:methanol_dcs}, we list the derivative couplings between CIS states 1 and 4. Note that correcting the derivative coupling with the ERFs introduced only a small change. To facilitate comparison, we also list in Table \ref{tab:methanol_dcs} the derivative couplings with only ETF corrections, and with both ETF and ERF corrections. For small molecules, the ETF correction is known to change the derivative coupling more dramatically\cite{fatehi:2011:dercouple}, but we see in Fig \ref{fig:methanol_dc} that the ERFs  appear have a smaller effect in this case. In the future  we will need to investigate the size of the ERF correction and ascertain exactly when an ERF correction will be essential because angular momentum conservation is paramount (as it was in Ref. \citenum{truhlar:2020:project_rot_nac}): 
\begin{table}[]
\centering
\caption{Derivative couplings (in \si{\per\bohr}) between the first and the fourth CIS states for methanol according to a def2-svp basis. The derivative couplings are computed in four possible ways: (a) no corrections made to the derivative couplings, (b) applying ERFs, and (c) applying ETFs, and finally (d) applying both ETFs and ERFs. For this molecule, applying ETFs changes the derivative coupling far more than applying ERFs. Future work will be necessary to identify the overall magnitude of this correction to the momentum rescaling direction.}
\label{tab:methanol_dcs}
\begin{tabular}{@{}lrrrrrrrrrrrrrrr@{}}
\toprule
Atom &
  \multicolumn{3}{c}{No Correction} &
  \multicolumn{1}{c}{} &
  \multicolumn{3}{c}{ERF only} &
  \multicolumn{1}{c}{} &
  \multicolumn{3}{c}{ETF only} &
  \multicolumn{1}{c}{} &
  \multicolumn{3}{c}{ETF and ERF} \\ \cmidrule(lr){2-4} \cmidrule(lr){6-8} \cmidrule(lr){10-12} \cmidrule(l){14-16} 
 &
  \multicolumn{1}{c}{$x$} &
  \multicolumn{1}{c}{$y$} &
  \multicolumn{1}{c}{$z$} &
  \multicolumn{1}{c}{} &
  \multicolumn{1}{c}{$x$} &
  \multicolumn{1}{c}{$y$} &
  \multicolumn{1}{c}{$z$} &
  \multicolumn{1}{l}{} &
  \multicolumn{1}{c}{$x$} &
  \multicolumn{1}{c}{$y$} &
  \multicolumn{1}{c}{$z$} &
  \multicolumn{1}{c}{} &
  \multicolumn{1}{c}{$x$} &
  \multicolumn{1}{c}{$y$} &
  \multicolumn{1}{c}{$z$} \\ \midrule
C    & 0.3447                  & -0.0043                 & 0.0003                  &                      & 0.3398                  & -0.0168                 & -0.0003                 &                      & 0.2287                  & -0.0671                 & 0.0000                  &                      & 0.2255                  & -0.0761                 & -0.0004                 \\
H    & 0.0142                  & -0.0329                 & 0.0158                  &                      & 0.0113                  & -0.0360                 & 0.0157                  &                      & 0.0484                  & -0.0332                 & 0.0131                  &                      & 0.0441                  & -0.0375                 & 0.0129                  \\
H    & 0.0136                  & -0.0326                 & -0.0146                 &                      & 0.0097                  & -0.0348                 & -0.0148                 &                      & 0.0489                  & -0.0326                 & -0.0119                 &                      & 0.0434                  & -0.0363                 & -0.0122                 \\
H    & -0.0531                 & 0.0352                  & 0.0005                  &                      & -0.0477                 & 0.0339                  & 0.0006                  &                      & -0.0818                 & 0.0350                  & 0.0005                  &                      & -0.0747                 & 0.0327                  & 0.0005                  \\
O    & -0.4608                 & -0.3230                 & -0.0040                 &                      & -0.4321                 & -0.3088                 & -0.0027                 &                      & -0.4668                 & -0.3545                 & -0.0032                 &                      & -0.4399                 & -0.3406                 & -0.0023                 \\
H    & 0.2575                  & 0.4496                  & 0.0020                  &                      & 0.2352                  & 0.4544                  & 0.0015                  &                      & 0.2226                  & 0.4524                  & 0.0016                  &                      & 0.2016                  & 0.4579                  & 0.0014                  \\ \bottomrule
\end{tabular}
\end{table}

\section{Discussion and Conclusions} \label{sec-conc}
Linear momentum and angular momentum are good quantum numbers for molecular systems, and thus
neither total translations nor total rotations can induce a nonadiabatic electron change of state.  That being said, problems with conservation arise within a surface hopping formalism because the algorithm completely ignores all electronic linear and angular momentum (which are assumed to be zero because all real-valued electronic states carry zero expectation value for linear and angular momentum). Thus, within a simple surface hopping formalism, it is perhaps not surprising that linear and angular momentum conservation is violated routinely. 

Here, we have reviewed how  electron translation factors are introduced into the surface hopping ansatz, and we have introduced electron rotation factors.  The former removes the translational components of the derivative couplings, and the latter removes the rotational components of the derivative coupling. These dressed derivative couplings have been constructed in a reasonably physical way, and they are size consistent: One can be sure that one will never produce a derivative coupling delocalized over two noninteracting systems.  If one uses the modified derivative couplings presented above (Eq. \ref{easy1final}), one can be absolutely certain that momentum rescaling will not break linear or angular momentum conservation.

At this point, one more word is appropriate regarding the problem of electronic linear and angular momentum within the surface hopping algorithm.  Although so far we have focused mostly on the rescaling direction after one hops, we have also noted in Eq. \ref{dressedGamm2} that the hopping rate itself is changed by the inclusion of electron translation factors.  To that end, our opinion is that the electron momentum problem will have a greater impact on the question  of the rescaling direction rather  the question of when to hop.  After all, in practice, the effect of translation and rotation insofar as driving an electronic transition is usually quite small.

Nevertheless,  if one really wanted to, one could in principle apply the ETF and ERF corrections above to the hopping rate as well.  However, one must be careful when doing so because  no one today calculates the derivative coupling in  Eq. \ref{hop2} when deciding if to hop\cite{tully:fssh}.  Instead,  the usual practice nowadays\cite{shs:1994:protons,Meek2014} is to evaluate the hopping rate numerically using the overlap of the electronic states at different times. For instance, if one is moving from $R(t)$ to $R(t + \Delta t)$, one calculates the overlap $U_{JK} = \left< \Phi_J(t) \middle| \Phi_K(t+ \Delta t) \right>$.  From this overlap, {\bf U}, one can evaluate the hopping probability for a time window $dt$ by taking the matrix logarithm of {\bf U}\cite{jain:2016:fast} (and compare with Eq. \ref{hop2}):
\begin{eqnarray}
\label{lasthop}
    \Gamma_{J\rightarrow K} =2 \mbox{Re} \left(\frac{ \log({\bf U})}{dt} \frac{\rho_{KJ}}{\rho_{JJ}}\right) 
\end{eqnarray}
In principle, one could add the ETF and ERF corrections above in Eqs. \ref{detf} and \ref{easy1final} to the hopping expression in Eq. \ref{lasthop}.   
In practice, however, we note that many MD codes today insist that at time zero, the total linear and angular momentum be zero and  this zeroing of the center of mass motion can solve many problems automatically, e.g.  the H-atom problem in Section \ref{sec-lin}.  More generally, if one runs a surface hopping calculation with nonzero nuclear linear and/or angular momentum, when calculating the hopping rate from Eq. \ref{hop2}, one usually translates and rotate the molecular geometries  between time steps so as to make sure that the two geometries are maximally aligned with the same center of mass. These standard rules of thumb (which have nothing to do with the theory described above) should partially mitigate the momentum conservation problem encountered when hopping.

Looking forward, Eq. \ref{easy1final} is incredibly easy to implement and should be broadly applied in the community.  That being said, a few interesting questions remain as of yet, unanswered, and will need further research. First, the present theory has relied on translational and rotational invariance of the total Hamiltonian (electronic plus nuclear). Thus, these results do not hold in the presence of an external electric or magnetic field: understanding how to properly evaluate the overall translational motion and angular momentum within such an external field is not obvious. Second, as mentioned above, our treatment above excluded the possibility of any spin-orbit coupling; we specifically did not treat the case where the electronic angular momentum of an eigenstate was nonzero, which will lead to further complications. For example, recent work\cite{bian:2023:angcons} has demonstrated that Berry forces of the form
\begin{align}
F_J^{A\alpha} =2 \hbar \mbox{Im} \left( \sum_{K B \beta} d_{JK}^{A\alpha} d_{KJ}^{B \beta} v_{B \beta} \right)
\end{align}
must be included in order for Born-Oppenheimer motion along  a single Kramer's adiabat to conserve angular momentum. Clearly, in such a case, one cannot make the crude approximation for the derivative coupling as in Eq. \ref{easy1final} and simply ignore all electronic and spin angular momentum. 
Third, the situation for photoemission of electrons is equally problematic; clearly, in such a case, one should not simply ignore the electronic momentum in the context of nonadiabatic dynamics.

In conclusion, as is well known, standard surface hopping algorithm ignores many nuances of nonadiabatic dynamics. In the present paper, we have focused on one such omission: the neglect of electronic and angular momentum.  Recovering a surface hopping algorithm that conserve momentum really boils down to finding a meaningful momentum rescaling   direction, and here we have presented Eq. \ref{easy1final}) as one reasonable option.  In the future, going beyond standard surface hopping and accounting for electronic momenta  appears to  be an extremely fertile area for theoretical development, especially given recent heightened attention to chiral induced spin selectivity\cite{naaman:2012:jpcl,naaman:2019:natrev,das:2022:cisstemp} and the need to understand the coupling of electronic and nuclear spins more broadly in spin-lattice relaxation\cite{cox:1997:hyperfine,zheng:2022:shspinlatt}.  As discussed in the Appendix, phase space approaches\cite{wu:2022:pssh,bian:2022:pssh} (e.g. PSSH) are extremely interesting candidates for solving these problems, but any and all other solutions must also be investigated. 



\section{Acknowledgments}
This material is based on the work supported by the National Science Foundation under Grant No. CHE-2102402. 

\appendix

\section{Geometries Used}
Here we list the $xyz$ co-ordinates (in \unit{\angstrom} ) for the methanol molecule studied in Sec. \ref{sec-res}.
\begin{table}[h]
\renewcommand{\tabcolsep}{0.35cm}
\begin{tabular}{@{}lrrr@{}}
Atom & \multicolumn{1}{c}{$x$} & \multicolumn{1}{c}{$y$} & \multicolumn{1}{c}{$z$} \\
C    & -0.6529987839           & 0.0229298748            & -0.0000329215           \\
H    & -0.9804779514           & 0.5600418428            & -0.9165130319           \\
H    & -0.9800978088           & 0.5649365905            & 0.9137178379            \\
H    & -1.1275275893           & -0.9798785707           & 0.0028970309            \\
O    & 0.7365265461            & -0.1333857115           & 0.0000189022            \\
H    & 1.1138836843            & 0.7844065804            & -0.0000555256          
\end{tabular}
\end{table}

\bigskip
\bigskip
\bigskip

\pagebreak
\section{Phase-Space Surface Hopping Interpretation of Electron Translation}
Interestingly, the result in Eq. \ref{longdetf} (which was derived above using electron translation factors) can be established within a different framework by working in coordinates where all coordinates are chosen relative to the position of one given nucleus (here, denoted ${\bf R}_A$). In such a case, we can imagine transforming from coordinates $\left({\bf r}_1, \cdots, {\bf r}_n, {\bf R}_1, \ldots, {\bf R}_n\right)$ to coordinates
$\left({\bf r'}_1, \cdots, {\bf r'}_n, {\bf R'}_1, \ldots, {\bf R'}_n\right)$
:
\begin{align}
\label{trans1}
{\bf R'}_B &\equiv {\bf R}_B \\
\label{trans2}
{\bf r'}_j &\equiv {\bf r}_j - {\bf R}_A
\end{align}
The corresponding relationships for the derivatives are:
\begin{align}
\label{trans3}
\frac{\partial}{\partial {\bf r}_j}
 &= 
 \frac{\partial}{\partial {\bf r'}_j} \\
 \label{trans4}
\frac{\partial}{\partial {\bf R}_B}
 &= 
 \frac{\partial}{\partial {\bf R'}_B}- \delta_{AB} \sum_j  \frac{\partial}{\partial {\bf r'}_j}
\end{align}

Simple algebra now shows that the total Hamiltonian
\begin{align}
    \hat{H}_{tot} &= \sum_B  \frac{\hat{\bf P}_B\cdot \hat{\bf P}_B}{2m_B} + \hat{H}_{el} \\
    \hat{H}_{el} & = \sum_i \frac{\hat{\bf p}_i\cdot \hat{\bf p_i}}{2m_e} + V\left(\hat{\bf r}_1, \cdots, \hat{\bf r}_n, \hat{\bf R}_1, \ldots, \hat{\bf R}_n\right)
\end{align}
now transforms into
\begin{align}
    \hat{H}_{tot} & = \sum_B  \frac{\hat{\bf P}'_B\cdot \hat{\bf P}'_B}{2m_B} + \hat{H}_{el}' \\
    \hat{H}_{el}' & = \sum_i \frac{\hat{\bf p}'_i\cdot \hat{\bf p}'_i}{2m_e} + V\left(\hat{\bf r}'_1, \cdots, \hat{\bf r}'_n, \hat{\bf R}'_1, \ldots, \hat{\bf R}'_n\right) 
+  \frac{1}{2 m_A} \sum_{i,j} \hat{\bf p}'_i\cdot \hat{\bf p}'_j 
    -\frac{\hat{\bf P}'_A}{m_A}\sum_j \hat{\bf p}'_j
    \label{newcorr}
\end{align}
The third and fourth terms in Eq. \ref{newcorr} are the corrections induced by the point transformation in Eqs. \ref{trans1}-\ref{trans2}.  
The third term in Eq. \ref{newcorr} is the analogous of the  mass polarization term  (but is different because we use electronic coordinates relative to one specific nucleus rather than the total center of mass); this term can easily be grouped into the electronic Hamiltonian and dealt with (but it is small). The fourth term in Eq. \ref{newcorr} is  more interesting. This term is responsible for electron translation coupled with nuclear translation, which cannot easily be grouped within the electronic Hamiltonian (if we want $\hat{H}_{el}$ to be a function of the $\Rb$'s only). However, if we make a semiclassical ansatz and replace the quantum operator $\hat{\bf P}_A$ in Eq. \ref{newcorr} with the classical coordinate ${\bf P}_A$, we can include this term in the electronic Hamiltonian (which now depends on both the $\Rb$'s and $\Pb$'s). Such ``phase-space'' electronic Hamiltonians have been explored in different contexts within surface hopping recently\cite{wu:2022:pssh,bian:2022:pssh}.
If we then treat the term $\left( \frac{{\bf P}'_A}{m_A}\sum_j {\bf p}'_j \right)$ perturbatively and, as above, consider the form of the Hamiltonian matrix element $H_{JK}$ that enters into Eq. \ref{hop} for the hopping rate, we will find the term $\frac{-i}{\hbar} \sum_{\mu, \nu, B, \alpha} 
D^{JK}_{\mu \nu} v_{B \alpha} \tilde{S}^{A \alpha}_{\nu \mu}$ for nucleus $A$ (which is the same correction as we found above in Eq. \ref{newJK}). Note, however, that in principle, if we wanted to reproduce the electron translation factors outlined above in Eq. \ref{detf}, we would need to perform such a different calculation for each nucleus $A$.

The bottom line is that when we want to consider hopping between electronic states semiclassically, and we want to ignore electronic momentum,
the ETF dressed derivative coupling in Eq. \ref{detf}
is a good first order approximation. That being said, a phase space approach  offers a more rigorous starting point for more accurate treatments that seek a more complete account of both electron and nuclear momentum; and  in the future, one should be able to use a single, rigorous  phase-space Hamiltonian to recover (angular and linear) momentum conservation rather than the set of nucleus specific Hamiltonians described above that simply reproduced our description of ETFs.  Most importantly, in the future, such phase space\cite{wu:2022:pssh,bian:2022:pssh} approaches may well be promising for treating the angular momentum problem where there is spin-orbit coupling and one cannot make the approximations described in Sections \ref{sec-ang} and \ref{sec-conc} above.

\section{Lagrange Multipliers As a Naive Approach to ETFs and ERFs} \label{tian}

In the main body of the text above, we have shown how to construct ETFs and ERFs heuristically.
Here, we will show that the final results (Eqs. 
\ref{dnot2} and \ref{easy1final}) are not the obvious expressions derived from Lagrange multipliers. For the sake of notational simplicity, we use bold symbols below to represent column vectors in $\mathbb{R}^3$ .

\subsection{Electron Translation Factors}
We begin with ETFs and we will show that using Lagrange multipliers leads to Eq. \ref{dnot2} above.
Given the derivative coupling vectors $\bm{d}^B$, one would like to apply small corrections that force $\sum_B\bm{d}^B = \bm{0}$. One may choose to modify $\bm{d}^B$ in the following way:
\begin{align}
    \tilde{\bm{d}}^B = \bm{d}^B + \sqrt{|\bm{d}^B|}\bm{x}^B
\end{align}
Mathematically, this corresponds to optimizing the function 
\begin{align}
    f(\bm{x})=\sum_B \zeta^B \bm{x}^{B\top}\bm{x}^{B}
\end{align}
subject to the constraint
\begin{align}
    \sum_B \left(\bm{d}^B + \sqrt{|\bm{d}^B|}\bm{x}^B\right) = \bm{0}\label{eq:l1_c1}
\end{align}
This optimization can be achieved with the following objective function with Lagrange multiplier $\bm{\lambda}$:
\begin{align}
    \mathcal{L} &= \frac{1}{2}\sum_B\zeta^B\bm{x}^{B\top}\bm{x}^{B} - \bm{\lambda}^\top\sum_B\left(\bm{d}^B + \sqrt{|\bm{d}^B|}\bm{x}^B\right)
\end{align}
Taking the derivative of $\mathcal{L}$ w.r.t. $\bm{x}^B$ leads to
\begin{align}
    \bm{x}^B = \frac{1}{\zeta^B}\sqrt{|\bm{d}^B|}\bm{\lambda}
\end{align}
Inserting this expression into the constraint Eq. \ref{eq:l1_c1} yields
\begin{align}
    &\sum_B \left(\bm{d}^B + \frac{|\bm{d}^B|}{\zeta^B}\bm{\lambda}\right) = \bm{0}\\
    \Rightarrow & \bm{\lambda} = -\frac{\sum_B\bm{d}^B}{\sum_B\frac{|\bm{d}^B|}{\zeta^B}}
\end{align}
Therefore, we have
\begin{align} 
    \tilde{\bm{d}}^B &= \bm{d}^B + \sqrt{|\bm{d}^B|}\bm{x}^B\\
    & = \bm{d}^B + \frac{|\bm{d}^B|}{\zeta^B}\bm{\lambda}\\
    & \label{detftian} =\bm{d}^B - \frac{|\bm{d}^B|/\zeta^B}{\sum_A|\bm{d}^A|/\zeta^A}\sum_A\bm{d}^A
\end{align}
which is identical to Eq. \ref{dnot2} when $\zeta^B = 1$. 

Note that Eq. \ref{detftian} is not equivalent to  Eq. \ref{detf} above.

\subsection{Electron Rotation Factors}
Next, we will show what the relevant minimization formula would be when deriving ERFs. As before, we will find  that the final answer is different  from the derived equation (Eq. \ref{easy1final}).
Given the nuclear coordinates $\bm{R}^B$ and the derivative coupling vectors $\bm{d}^B$, one may again choose to correct $\bm{d}^B$ by $\tilde{\bm{d}}^B = \bm{d}^B + \sqrt{|\bm{d}^B|}\bm{x}^B$ and to minimize $\sum_B \bm{x}^{B\top}\bm{x}^{B}$, but this time subject to the constraints that
\begin{align}
    \sum_{B}\bm{R}^B \times \left(\bm{d}^B + \sqrt{|\bm{d}^B|}\bm{x}^B\right) &= \bm{0}\label{eq:l2_c1} \\ 
    \sum_B \sqrt{|\bm{d}^B|}\bm{x}^B &= \bm{0} \label{eq:l2_c2}
\end{align}
where ``$\times$" is the vector cross product symbol. For any two column vectors $\bm{A}$ and $\bm{B}$ in $\mathbb{R}^3$, the cross-product is defined by
\begin{align}
    \left(\bm{A}\times\bm{B}\right)^\alpha = \sum_{\beta\gamma}\epsilon_{\alpha\beta\gamma}A^\beta B^\gamma
\end{align}

Similar to the case of ETFs, the optimization can be achieved with the following objective function, now with Lagrange multipliers $\bm{\lambda}_1$ and $\bm{\lambda}_2$:
\begin{align}
    \mathcal{L} = \frac{1}{2}\sum_{B}\bm{x}^{B\top}\bm{x}^{B} - \bm{\lambda}_1^{\top}\sum_B\left(\bm{R}^B \times \bm{d}^B + \sqrt{|\bm{d}^B|}\bm{R}^B \times \bm{x}^B\right)-\bm{\lambda}_2^{\top}\sum_B \sqrt{|\bm{d}^B|}\bm{x}^B
\end{align}
Again, taking the derivative of $\mathcal{L}$ w.r.t. $\bm{x}^B$ leads to
\begin{align}
    \bm{x}^B = \sqrt{|\bm{d}^B|}\bm{\lambda}_1\times\bm{R}^B + \sqrt{|\bm{d}^B|}\bm{\lambda}_2
\end{align}
To solve for $\bm{\lambda}_1$ and $\bm{\lambda}_2$, we substitute $\bm{x}^B$ into the constraints Eq. \ref{eq:l2_c1} and Eq. \ref{eq:l2_c2}:
\begin{align}
    \sum_B \bm{R}^B\times\bm{d}^B + \sum_B|\bm{d}^B|\bm{R}^B\times\left(\bm{\lambda}_1\times\bm{R}^B\right) + \sum_B|\bm{d}^B|\bm{R}^B\times\bm{\lambda}_2 &= \bm{0}\label{eq:l2_d1} \\
    \bm{\lambda}_1\times\sum_B|\bm{d}^B|\bm{R}^B + \bm{\lambda}_2\sum_B|\bm{d}^B| &= \bm{0}\label{eq:l2_d2}
\end{align}
From Eq. \ref{eq:l2_d2}
\begin{align}
    \bm{\lambda}_2 = -\frac{\bm{\lambda}_1\times\sum_B|\bm{d}^B|\bm{R}^B}{\sum_B|\bm{d}^B|}\label{eq:l2_d3}
\end{align}
Let
\begin{align}
    \bm{W}^B &=\sqrt{|\bm{d}^B|}\bm{R}^B\\
    \bm{W} &= \frac{\sum_B|\bm{d}^B|\bm{R}^B}{\sqrt{\sum_B|\bm{d}^B|}}
\end{align}
Inserting Eq. \ref{eq:l2_d3} into Eq. \ref{eq:l2_d1} yields
\begin{align}
    \bm{W}\times\left(\bm{\lambda}_1\times\bm{W}\right)-\sum_B\bm{W}^B\times\left(\bm{\lambda}_1\times\bm{W}^B\right) = \sum_B \bm{R}^B\times\bm{d}^B\label{eq:l2_d4}
\end{align}
Let us define the 3 by 3 matrix $\mathcal{M}$ 
\begin{align}
    \mathcal{M} = \left(\bm{W}^\top\bm{W}-\sum_B\bm{W}^{B\top}\bm{W}^B\right)\mathcal{I} - \left(\bm{W}\bm{W}^\top - \sum_B\bm{W}^B\bm{W}^{B\top}\right)
\end{align}
where $\mathcal{I}$ is a 3 by 3 identity matrix. The solution to Eq. \ref{eq:l2_d4} is
\begin{align}
    \bm{\lambda}_1 = \mathcal{M}^{-1}\left(\sum_B \bm{R}^B\times\bm{d}^B\right)
\end{align}
and the correction to $\bm{d}^B$ is 
\begin{align}
    \tilde{\bm{d}}^B &=\bm{d}^B + \sqrt{|\bm{d}^B|}\bm{x}^B\\
    &=\bm{d}^B + |\bm{d}^B|(\bm{\lambda}_1\times\bm{R}^B+\bm{\lambda}_2)\\
    &=\bm{d}^B + |\bm{d}^B|\bm{\lambda}_1\times\left(\bm{R}^B - \frac{\sum_A|\bm{d}^A|\bm{R}^A}{\sum_A|\bm{d}^A|}\right)\\
    &=\bm{d}^B + |\bm{d}^B|\left[\mathcal{M}^{-1}\left(\sum_A \bm{R}^A\times\bm{d}^A\right)\right]\times\left(\bm{R}^B - \frac{\sum_A|\bm{d}^A|\bm{R}^A}{\sum_A|\bm{d}^A|}\right)\label{eq:l2_final}
\end{align}
{Note that, when one enforces a ``centroid'' condition (i.e. $\sum_B\bm{R}^B=0.$ ) as in Ref.~\citenum{truhlar:2020:project_rot_nac}, Eq.~\ref{eq:l2_final} can be reduced exactly to the result in Eq.~11 in Ref.~\citenum{truhlar:2020:project_rot_nac}, 
provided we do not weight the correction term $\bm{x}^B$ by  $\sqrt{\bm{d}^B}$ (and thus sacrifice size consistency).} Either way, we emphasize that Eq. \ref{eq:l2_final} is not
equivalent to Eq. \ref{easy1final}.

As a sanity check, one can easily show that these expressions satisfy Eqs. \ref{eq:l2_c1} and \ref{eq:l2_c2} as follows:
For checking Eq. \ref{eq:l2_c1},
\begin{align}
    &\sum_B \bm{R}^B \times \tilde{\bm{d}}^B\\
    =&\sum_B\bm{R}^B\times\bm{d}^B + \sum_B|\bm{d}^B|\bm{R}^B\times(\bm{\lambda}_1\times\bm{R}^B)-\sum_B|\bm{d}^B|\bm{R}^B\times\left(\bm{\lambda}_1\times\frac{\sum_A|\bm{d}^A|\bm{R}^A}{\sum_A|\bm{d}^A|}\right)\\
    =&\sum_B\bm{R}^B\times\bm{d}^B + \sum_B\bm{W}^B\times(\bm{\lambda}_1\times\bm{W}^B)-\bm{W}\times(\bm{\lambda}_1\times\bm{W})\\
    =&\sum_B\bm{R}^B\times\bm{d}^B - \mathcal{M}\bm{\lambda}_1\\
    =&\sum_B\bm{R}^B\times\bm{d}^B-\sum_A\bm{R}^A\times\bm{d}^A\\
    =&\bm{0}
\end{align}
As for checking Eq. \ref{eq:l2_c2},
\begin{align}
    &\sum_B |\bm{d}^B|\left[\mathcal{M}^{-1}\left(\sum_A \bm{R}^A\times\bm{d}^A\right)\right]\times\left(\bm{R}^B - \frac{\sum_A|\bm{d}^A|\bm{R}^A}{\sum_A|\bm{d}^A|}\right)\\
    =&\left[\mathcal{M}^{-1}\left(\sum_A \bm{R}^A\times\bm{d}^A\right)\right]\times\left(\sum_B|\bm{d}^B|\bm{R}^B - \sum_B|\bm{d}^B|\frac{\sum_A|\bm{d}^A|\bm{R}^A}{\sum_A|\bm{d}^A|}\right)\\
    =&\left[\mathcal{M}^{-1}\left(\sum_A \bm{R}^A\times\bm{d}^A\right)\right]\times\left(\sum_B|\bm{d}^B|\bm{R}^B-\sum_A|\bm{d}^A|\bm{R}^A\right)\\
    =&\bm{0}
\end{align}

\bibliography{finalbib}

\begin{thebibliography}{54}%
\makeatletter
\providecommand \@ifxundefined [1]{%
 \@ifx{#1\undefined}
}%
\providecommand \@ifnum [1]{%
 \ifnum #1\expandafter \@firstoftwo
 \else \expandafter \@secondoftwo
 \fi
}%
\providecommand \@ifx [1]{%
 \ifx #1\expandafter \@firstoftwo
 \else \expandafter \@secondoftwo
 \fi
}%
\providecommand \natexlab [1]{#1}%
\providecommand \enquote  [1]{``#1''}%
\providecommand \bibnamefont  [1]{#1}%
\providecommand \bibfnamefont [1]{#1}%
\providecommand \citenamefont [1]{#1}%
\providecommand \href@noop [0]{\@secondoftwo}%
\providecommand \href [0]{\begingroup \@sanitize@url \@href}%
\providecommand \@href[1]{\@@startlink{#1}\@@href}%
\providecommand \@@href[1]{\endgroup#1\@@endlink}%
\providecommand \@sanitize@url [0]{\catcode `\\12\catcode `\$12\catcode
  `\&12\catcode `\#12\catcode `\^12\catcode `\_12\catcode `\%12\relax}%
\providecommand \@@startlink[1]{}%
\providecommand \@@endlink[0]{}%
\providecommand \url  [0]{\begingroup\@sanitize@url \@url }%
\providecommand \@url [1]{\endgroup\@href {#1}{\urlprefix }}%
\providecommand \urlprefix  [0]{URL }%
\providecommand \Eprint [0]{\href }%
\providecommand \doibase [0]{https://doi.org/}%
\providecommand \selectlanguage [0]{\@gobble}%
\providecommand \bibinfo  [0]{\@secondoftwo}%
\providecommand \bibfield  [0]{\@secondoftwo}%
\providecommand \translation [1]{[#1]}%
\providecommand \BibitemOpen [0]{}%
\providecommand \bibitemStop [0]{}%
\providecommand \bibitemNoStop [0]{.\EOS\space}%
\providecommand \EOS [0]{\spacefactor3000\relax}%
\providecommand \BibitemShut  [1]{\csname bibitem#1\endcsname}%
\let\auto@bib@innerbib\@empty
\bibitem [{\citenamefont {Tully}(1990)}]{tully:fssh}%
  \BibitemOpen
  \bibfield  {author} {\bibinfo {author} {\bibfnamefont {J.~C.}\ \bibnamefont
  {Tully}},\ }\bibfield  {title} {\enquote {\bibinfo {title} {Molecular
  dynamics with electronic transitions},}\ }\href@noop {} {\bibfield  {journal}
  {\bibinfo  {journal} {Journal of Chemical Physics}\ }\textbf {\bibinfo
  {volume} {93}},\ \bibinfo {pages} {1061--1071} (\bibinfo {year}
  {1990})}\BibitemShut {NoStop}%
\bibitem [{\citenamefont {Barbatti}(2011)}]{barbatti:2011:review}%
  \BibitemOpen
  \bibfield  {author} {\bibinfo {author} {\bibfnamefont {M.}~\bibnamefont
  {Barbatti}},\ }\bibfield  {title} {\enquote {\bibinfo {title} {Nonadiabatic
  dynamics with trajectory surface hopping method},}\ }\href@noop {} {\bibfield
   {journal} {\bibinfo  {journal} {Wiley Interdisciplinary Reviews:
  Computational Molecular Science}\ }\textbf {\bibinfo {volume} {1}},\ \bibinfo
  {pages} {620--633} (\bibinfo {year} {2011})}\BibitemShut {NoStop}%
\bibitem [{\citenamefont {Nelson}\ \emph {et~al.}(2014)\citenamefont {Nelson},
  \citenamefont {Fernandez-Alberti}, \citenamefont {Roitberg},\ and\
  \citenamefont {Tretiak}}]{tretiak:2014:acr}%
  \BibitemOpen
  \bibfield  {author} {\bibinfo {author} {\bibfnamefont {T.}~\bibnamefont
  {Nelson}}, \bibinfo {author} {\bibfnamefont {S.}~\bibnamefont
  {Fernandez-Alberti}}, \bibinfo {author} {\bibfnamefont {A.~E.}\ \bibnamefont
  {Roitberg}},\ and\ \bibinfo {author} {\bibfnamefont {S.}~\bibnamefont
  {Tretiak}},\ }\bibfield  {title} {\enquote {\bibinfo {title} {Nonadiabatic
  excited-state molecular dynamics: Modeling photophysics in organic conjugated
  materials},}\ }\href@noop {} {\bibfield  {journal} {\bibinfo  {journal}
  {Accounts of Chemical Research}\ }\textbf {\bibinfo {volume} {47}},\ \bibinfo
  {pages} {1155--1164} (\bibinfo {year} {2014})}\BibitemShut {NoStop}%
\bibitem [{\citenamefont {Parandekar}\ and\ \citenamefont
  {Tully}(2005)}]{tully:2005:detailedbalance}%
  \BibitemOpen
  \bibfield  {author} {\bibinfo {author} {\bibfnamefont {P.~V.}\ \bibnamefont
  {Parandekar}}\ and\ \bibinfo {author} {\bibfnamefont {J.~C.}\ \bibnamefont
  {Tully}},\ }\bibfield  {title} {\enquote {\bibinfo {title} {Mixed
  quantum-classical equilibrium},}\ }\href@noop {} {\bibfield  {journal}
  {\bibinfo  {journal} {Journal of Chemical Physics}\ }\textbf {\bibinfo
  {volume} {122}},\ \bibinfo {pages} {094102} (\bibinfo {year}
  {2005})}\BibitemShut {NoStop}%
\bibitem [{\citenamefont {Schmidt}, \citenamefont {Parandekar},\ and\
  \citenamefont {Tully}(2008)}]{tully:2008:detailedbalance}%
  \BibitemOpen
  \bibfield  {author} {\bibinfo {author} {\bibfnamefont {J.~R.}\ \bibnamefont
  {Schmidt}}, \bibinfo {author} {\bibfnamefont {P.~V.}\ \bibnamefont
  {Parandekar}},\ and\ \bibinfo {author} {\bibfnamefont {J.~C.}\ \bibnamefont
  {Tully}},\ }\bibfield  {title} {\enquote {\bibinfo {title} {Mixed
  quantum-classical equilibrium: Surface hopping},}\ }\href@noop {} {\bibfield
  {journal} {\bibinfo  {journal} {Journal of Chemical Physics}\ }\textbf
  {\bibinfo {volume} {129}},\ \bibinfo {pages} {044104} (\bibinfo {year}
  {2008})}\BibitemShut {NoStop}%
\bibitem [{\citenamefont {Send}\ and\ \citenamefont
  {Furche}(2010)}]{furche:2010:nact}%
  \BibitemOpen
  \bibfield  {author} {\bibinfo {author} {\bibfnamefont {R.}~\bibnamefont
  {Send}}\ and\ \bibinfo {author} {\bibfnamefont {F.}~\bibnamefont {Furche}},\
  }\bibfield  {title} {\enquote {\bibinfo {title} {First-order nonadiabatic
  couplings from time-dependent hybrid density functional response theory:
  Consistent formalism, implementation, and performance},}\ }\href@noop {}
  {\bibfield  {journal} {\bibinfo  {journal} {Journal of Chemical Physics}\
  }\textbf {\bibinfo {volume} {132}},\ \bibinfo {pages} {044107} (\bibinfo
  {year} {2010})}\BibitemShut {NoStop}%
\bibitem [{\citenamefont {Kutzelnigg}(2007)}]{kutzelnigg:2007:mp}%
  \BibitemOpen
  \bibfield  {author} {\bibinfo {author} {\bibfnamefont {W.}~\bibnamefont
  {Kutzelnigg}},\ }\bibfield  {title} {\enquote {\bibinfo {title} {Which masses
  are vibrating or rotating in a molecule?}}\ }\href
  {https://doi.org/10.1080/00268970701604671} {\bibfield  {journal} {\bibinfo
  {journal} {Molecular Physics}\ }\textbf {\bibinfo {volume} {105}},\ \bibinfo
  {pages} {2627--2647} (\bibinfo {year} {2007})},\ \Eprint
  {https://arxiv.org/abs/https://doi.org/10.1080/00268970701604671}
  {https://doi.org/10.1080/00268970701604671} \BibitemShut {NoStop}%
\bibitem [{\citenamefont
  {Patchkovskii}(2012)}]{patchkovskii:2012:jcp:electronic_current}%
  \BibitemOpen
  \bibfield  {author} {\bibinfo {author} {\bibfnamefont {S.}~\bibnamefont
  {Patchkovskii}},\ }\bibfield  {title} {\enquote {\bibinfo {title} {Electronic
  currents and born-oppenheimer molecular dynamics},}\ }\href@noop {}
  {\bibfield  {journal} {\bibinfo  {journal} {The Journal of Chemical Physics}\
  }\textbf {\bibinfo {volume} {137}},\ \bibinfo {pages} {084109} (\bibinfo
  {year} {2012})}\BibitemShut {NoStop}%
\bibitem [{\citenamefont {Thachuk}, \citenamefont {Ivanov},\ and\ \citenamefont
  {Wardlaw}(1998)}]{wardlaw:1998:momentum_sh}%
  \BibitemOpen
  \bibfield  {author} {\bibinfo {author} {\bibfnamefont {M.}~\bibnamefont
  {Thachuk}}, \bibinfo {author} {\bibfnamefont {M.~Y.}\ \bibnamefont
  {Ivanov}},\ and\ \bibinfo {author} {\bibfnamefont {D.~M.}\ \bibnamefont
  {Wardlaw}},\ }\bibfield  {title} {\enquote {\bibinfo {title} {A semiclassical
  approach to intense-field above-threshold dissociation in the long wavelength
  limit. ii. conservation principles and coherence in surface hopping},}\
  }\href {https://doi.org/10.1063/1.477197} {\bibfield  {journal} {\bibinfo
  {journal} {The Journal of Chemical Physics}\ }\textbf {\bibinfo {volume}
  {109}},\ \bibinfo {pages} {5747--5760} (\bibinfo {year} {1998})},\ \Eprint
  {https://arxiv.org/abs/https://doi.org/10.1063/1.477197}
  {https://doi.org/10.1063/1.477197} \BibitemShut {NoStop}%
\bibitem [{\citenamefont {Drukker}(1999)}]{drukker:1999:review_fssh}%
  \BibitemOpen
  \bibfield  {author} {\bibinfo {author} {\bibfnamefont {K.}~\bibnamefont
  {Drukker}},\ }\bibfield  {title} {\enquote {\bibinfo {title} {Basics of
  surface hopping in mixed quantum/classical simulations},}\ }\href
  {https://doi.org/https://doi.org/10.1006/jcph.1999.6287} {\bibfield
  {journal} {\bibinfo  {journal} {Journal of Computational Physics}\ }\textbf
  {\bibinfo {volume} {153}},\ \bibinfo {pages} {225--272} (\bibinfo {year}
  {1999})}\BibitemShut {NoStop}%
\bibitem [{\citenamefont {Fatehi}\ \emph {et~al.}(2011)\citenamefont {Fatehi},
  \citenamefont {Alguire}, \citenamefont {Shao},\ and\ \citenamefont
  {Subotnik}}]{fatehi:2011:dercouple}%
  \BibitemOpen
  \bibfield  {author} {\bibinfo {author} {\bibfnamefont {S.}~\bibnamefont
  {Fatehi}}, \bibinfo {author} {\bibfnamefont {E.}~\bibnamefont {Alguire}},
  \bibinfo {author} {\bibfnamefont {Y.}~\bibnamefont {Shao}},\ and\ \bibinfo
  {author} {\bibfnamefont {J.~E.}\ \bibnamefont {Subotnik}},\ }\bibfield
  {title} {\enquote {\bibinfo {title} {Analytical derivative couplings between
  configuration interaction singles states with built-in translation factors
  for translational invariance},}\ }\href@noop {} {\bibfield  {journal}
  {\bibinfo  {journal} {Journal of Chemical Physics}\ }\textbf {\bibinfo
  {volume} {135}},\ \bibinfo {pages} {234105} (\bibinfo {year}
  {2011})}\BibitemShut {NoStop}%
\bibitem [{\citenamefont {Fatehi}\ and\ \citenamefont
  {Subotnik}(2012)}]{fatehi:2012:dercouple}%
  \BibitemOpen
  \bibfield  {author} {\bibinfo {author} {\bibfnamefont {S.}~\bibnamefont
  {Fatehi}}\ and\ \bibinfo {author} {\bibfnamefont {J.~E.}\ \bibnamefont
  {Subotnik}},\ }\bibfield  {title} {\enquote {\bibinfo {title} {Derivative
  couplings with built-in electron-translation factors: Application to
  benzene},}\ }\href@noop {} {\bibfield  {journal} {\bibinfo  {journal}
  {Journal of Physical Chemistry Letters}\ }\textbf {\bibinfo {volume} {3}},\
  \bibinfo {pages} {2039--2043} (\bibinfo {year} {2012})}\BibitemShut {NoStop}%
\bibitem [{\citenamefont {Shu}\ \emph {et~al.}(2020)\citenamefont {Shu},
  \citenamefont {Zhang}, \citenamefont {Varga}, \citenamefont {Parker},
  \citenamefont {Kanchanakungwankul}, \citenamefont {Sun},\ and\ \citenamefont
  {Truhlar}}]{truhlar:2020:project_rot_nac}%
  \BibitemOpen
  \bibfield  {author} {\bibinfo {author} {\bibfnamefont {Y.}~\bibnamefont
  {Shu}}, \bibinfo {author} {\bibfnamefont {L.}~\bibnamefont {Zhang}}, \bibinfo
  {author} {\bibfnamefont {Z.}~\bibnamefont {Varga}}, \bibinfo {author}
  {\bibfnamefont {K.~A.}\ \bibnamefont {Parker}}, \bibinfo {author}
  {\bibfnamefont {S.}~\bibnamefont {Kanchanakungwankul}}, \bibinfo {author}
  {\bibfnamefont {S.}~\bibnamefont {Sun}},\ and\ \bibinfo {author}
  {\bibfnamefont {D.~G.}\ \bibnamefont {Truhlar}},\ }\bibfield  {title}
  {\enquote {\bibinfo {title} {Conservation of angular momentum in direct
  nonadiabatic dynamics},}\ }\href@noop {} {\bibfield  {journal} {\bibinfo
  {journal} {The Journal of Physical Chemistry Letters}\ }\textbf {\bibinfo
  {volume} {11}},\ \bibinfo {pages} {1135--1140} (\bibinfo {year}
  {2020})}\BibitemShut {NoStop}%
\bibitem [{\citenamefont {Mead}\ and\ \citenamefont
  {Moscowitz}(1967)}]{mead_moscowitz:1967:dipole_vs_length}%
  \BibitemOpen
  \bibfield  {author} {\bibinfo {author} {\bibfnamefont {C.~A.}\ \bibnamefont
  {Mead}}\ and\ \bibinfo {author} {\bibfnamefont {A.}~\bibnamefont
  {Moscowitz}},\ }\bibfield  {title} {\enquote {\bibinfo {title} {Dipole length
  versus dipole velocity in the calculation of infrared intensities with
  {B}orn-{O}ppenheimer wave functions},}\ }\href@noop {} {\bibfield  {journal}
  {\bibinfo  {journal} {International Journal of Quantum Chemistry}\ }\textbf
  {\bibinfo {volume} {1}},\ \bibinfo {pages} {243} (\bibinfo {year}
  {1967})}\BibitemShut {NoStop}%
\bibitem [{\citenamefont {Nafie}(1983)}]{nafie:1983:jcp:elcurrent}%
  \BibitemOpen
  \bibfield  {author} {\bibinfo {author} {\bibfnamefont {L.~A.}\ \bibnamefont
  {Nafie}},\ }\bibfield  {title} {\enquote {\bibinfo {title} {{Adiabatic
  molecular properties beyond the Born–Oppenheimer approximation. Complete
  adiabatic wave functions and vibrationally induced electronic current
  density}},}\ }\href {https://doi.org/10.1063/1.445588} {\bibfield  {journal}
  {\bibinfo  {journal} {Journal of Chemical Physics}\ }\textbf {\bibinfo
  {volume} {79}},\ \bibinfo {pages} {4950--4957} (\bibinfo {year} {1983})},\
  \Eprint
  {https://arxiv.org/abs/https://pubs.aip.org/aip/jcp/article-pdf/79/10/4950/11239163/4950\_1\_online.pdf}
  {https://pubs.aip.org/aip/jcp/article-pdf/79/10/4950/11239163/4950\_1\_online.pdf}
  \BibitemShut {NoStop}%
\bibitem [{\citenamefont {Muller}\ and\ \citenamefont
  {Stock}(1997)}]{stock:1997:surfacehop}%
  \BibitemOpen
  \bibfield  {author} {\bibinfo {author} {\bibfnamefont {U.}~\bibnamefont
  {Muller}}\ and\ \bibinfo {author} {\bibfnamefont {G.}~\bibnamefont {Stock}},\
  }\bibfield  {title} {\enquote {\bibinfo {title} {Surface-hopping modeling of
  photoinduced relaxation dynamics on coupled potential-energy surfaces},}\
  }\href@noop {} {\bibfield  {journal} {\bibinfo  {journal} {Journal of
  Chemical Physics}\ }\textbf {\bibinfo {volume} {107}},\ \bibinfo {pages}
  {6230} (\bibinfo {year} {1997})}\BibitemShut {NoStop}%
\bibitem [{\citenamefont {Kelly}\ and\ \citenamefont
  {Markland}(2013)}]{markland:2013:surfacehop_marcus}%
  \BibitemOpen
  \bibfield  {author} {\bibinfo {author} {\bibfnamefont {A.}~\bibnamefont
  {Kelly}}\ and\ \bibinfo {author} {\bibfnamefont {T.~E.}\ \bibnamefont
  {Markland}},\ }\bibfield  {title} {\enquote {\bibinfo {title} {Efficient and
  accurate surface hopping for long time nonadiabatic quantum dynamics},}\
  }\href@noop {} {\bibfield  {journal} {\bibinfo  {journal} {Journal of
  Chemical Physics}\ }\textbf {\bibinfo {volume} {139}},\ \bibinfo {pages}
  {014104} (\bibinfo {year} {2013})}\BibitemShut {NoStop}%
\bibitem [{\citenamefont {Hazra}, \citenamefont {Soudakov},\ and\ \citenamefont
  {Hammes-Schiffer}(2010)}]{shs:2010:pcet_fssh}%
  \BibitemOpen
  \bibfield  {author} {\bibinfo {author} {\bibfnamefont {A.}~\bibnamefont
  {Hazra}}, \bibinfo {author} {\bibfnamefont {A.~V.}\ \bibnamefont
  {Soudakov}},\ and\ \bibinfo {author} {\bibfnamefont {S.}~\bibnamefont
  {Hammes-Schiffer}},\ }\bibfield  {title} {\enquote {\bibinfo {title} {Role of
  solvent dynamics in ultrafast photoinduced proton-coupled electron transfer
  reactions in solution},}\ }\href@noop {} {\bibfield  {journal} {\bibinfo
  {journal} {Journal of Physical Chemistry B}\ }\textbf {\bibinfo {volume}
  {114}},\ \bibinfo {pages} {12319--12332} (\bibinfo {year}
  {2010})}\BibitemShut {NoStop}%
\bibitem [{\citenamefont {Schwerdtfeger}, \citenamefont {Soudackov},\ and\
  \citenamefont {Hammes-Schiffer}(2014)}]{shs:2014:marcustheory_fssh}%
  \BibitemOpen
  \bibfield  {author} {\bibinfo {author} {\bibfnamefont {C.~A.}\ \bibnamefont
  {Schwerdtfeger}}, \bibinfo {author} {\bibfnamefont {A.~V.}\ \bibnamefont
  {Soudackov}},\ and\ \bibinfo {author} {\bibfnamefont {S.}~\bibnamefont
  {Hammes-Schiffer}},\ }\bibfield  {title} {\enquote {\bibinfo {title}
  {Nonadiabatic dynamics of electron transfer in solution: Explicit and
  implicit solvent treatments that include multiple relaxation time scales},}\
  }\href@noop {} {\bibfield  {journal} {\bibinfo  {journal} {Journal of
  Chemical Physics}\ }\textbf {\bibinfo {volume} {140}},\ \bibinfo {pages}
  {034113} (\bibinfo {year} {2014})}\BibitemShut {NoStop}%
\bibitem [{\citenamefont {Landry}\ and\ \citenamefont
  {Subotnik}(2011)}]{landry:2011:marcus_fssh}%
  \BibitemOpen
  \bibfield  {author} {\bibinfo {author} {\bibfnamefont {B.~R.}\ \bibnamefont
  {Landry}}\ and\ \bibinfo {author} {\bibfnamefont {J.~E.}\ \bibnamefont
  {Subotnik}},\ }\bibfield  {title} {\enquote {\bibinfo {title} {Standard
  surface hopping predicts incorrect scaling for marcus’ golden-rule rate:
  The decoherence problem cannot be ignored},}\ }\href@noop {} {\bibfield
  {journal} {\bibinfo  {journal} {Journal of Chemical Physics}\ }\textbf
  {\bibinfo {volume} {135}},\ \bibinfo {pages} {191101} (\bibinfo {year}
  {2011})}\BibitemShut {NoStop}%
\bibitem [{\citenamefont {Fuji}\ \emph {et~al.}(2010)\citenamefont {Fuji},
  \citenamefont {Suzuki}, \citenamefont {Horio}, \citenamefont {Suzuki},
  \citenamefont {Mitrić}, \citenamefont {Werner},\ and\ \citenamefont
  {Bonači\'c-Koutecký}}]{vbk:2010:furan}%
  \BibitemOpen
  \bibfield  {author} {\bibinfo {author} {\bibfnamefont {T.}~\bibnamefont
  {Fuji}}, \bibinfo {author} {\bibfnamefont {Y.-I.}\ \bibnamefont {Suzuki}},
  \bibinfo {author} {\bibfnamefont {T.}~\bibnamefont {Horio}}, \bibinfo
  {author} {\bibfnamefont {T.}~\bibnamefont {Suzuki}}, \bibinfo {author}
  {\bibfnamefont {R.}~\bibnamefont {Mitrić}}, \bibinfo {author} {\bibfnamefont
  {U.}~\bibnamefont {Werner}},\ and\ \bibinfo {author} {\bibfnamefont
  {V.}~\bibnamefont {Bonači\'c-Koutecký}},\ }\bibfield  {title} {\enquote
  {\bibinfo {title} {Ultrafast photodynamics of furan},}\ }\href@noop {}
  {\bibfield  {journal} {\bibinfo  {journal} {Journal of Chemical Physics}\
  }\textbf {\bibinfo {volume} {133}},\ \bibinfo {pages} {234303} (\bibinfo
  {year} {2010})}\BibitemShut {NoStop}%
\bibitem [{\citenamefont {Subotnik}, \citenamefont {Ouyang},\ and\
  \citenamefont {Landry}(2013)}]{subotnik:2013:qcle_fssh_derive}%
  \BibitemOpen
  \bibfield  {author} {\bibinfo {author} {\bibfnamefont {J.~E.}\ \bibnamefont
  {Subotnik}}, \bibinfo {author} {\bibfnamefont {W.}~\bibnamefont {Ouyang}},\
  and\ \bibinfo {author} {\bibfnamefont {B.~R.}\ \bibnamefont {Landry}},\
  }\bibfield  {title} {\enquote {\bibinfo {title} {Can we derive tully's
  surface-hopping algorithm from the semiclassical quantum liouville equation:
  Almost, but only with decoherence},}\ }\href@noop {} {\bibfield  {journal}
  {\bibinfo  {journal} {Journal of Chemical Physics}\ }\textbf {\bibinfo
  {volume} {139}},\ \bibinfo {pages} {214107} (\bibinfo {year}
  {2013})}\BibitemShut {NoStop}%
\bibitem [{\citenamefont {Kapral}(2016)}]{kapral:2016:chemphys_fssh}%
  \BibitemOpen
  \bibfield  {author} {\bibinfo {author} {\bibfnamefont {R.}~\bibnamefont
  {Kapral}},\ }\bibfield  {title} {\enquote {\bibinfo {title} {Surface hopping
  from the perspective of quantum–classical liouville dynamics},}\ }\href
  {https://doi.org/https://doi.org/10.1016/j.chemphys.2016.05.016} {\bibfield
  {journal} {\bibinfo  {journal} {Chemical Physics}\ }\textbf {\bibinfo
  {volume} {481}},\ \bibinfo {pages} {77 -- 83} (\bibinfo {year} {2016})},\
  \bibinfo {note} {quantum Dynamics and Femtosecond Spectroscopy dedicated to
  Prof. Vladimir Y. Chernyak on the occasion of his 60th birthday}\BibitemShut
  {NoStop}%
\bibitem [{\citenamefont {Landry}, \citenamefont {Falk},\ and\ \citenamefont
  {Subotnik}(2013)}]{landry:2013:electronicproperties}%
  \BibitemOpen
  \bibfield  {author} {\bibinfo {author} {\bibfnamefont {B.~R.}\ \bibnamefont
  {Landry}}, \bibinfo {author} {\bibfnamefont {M.~J.}\ \bibnamefont {Falk}},\
  and\ \bibinfo {author} {\bibfnamefont {J.~E.}\ \bibnamefont {Subotnik}},\
  }\bibfield  {title} {\enquote {\bibinfo {title} {Communication: The correct
  interpretation of surface hopping trajectories: How to calculate electronic
  properties},}\ }\href@noop {} {\bibfield  {journal} {\bibinfo  {journal}
  {Journal of Chemical Physics}\ }\textbf {\bibinfo {volume} {139}},\ \bibinfo
  {pages} {211101} (\bibinfo {year} {2013})}\BibitemShut {NoStop}%
\bibitem [{\citenamefont {Yarkony}(1989)}]{yarkony:1989:jcp_emailme_ang}%
  \BibitemOpen
  \bibfield  {author} {\bibinfo {author} {\bibfnamefont {D.~R.}\ \bibnamefont
  {Yarkony}},\ }\bibfield  {title} {\enquote {\bibinfo {title} {Nonadiabatic
  effects in the vicinity of multiple surface crossings. evaluation of
  derivative couplings with respect to rotational and internal degrees of
  freedom. application to the charge transfer reaction h++ no→ h+ no+},}\
  }\href@noop {} {\bibfield  {journal} {\bibinfo  {journal} {Journal of
  Chemical Physics}\ }\textbf {\bibinfo {volume} {90}},\ \bibinfo {pages}
  {1657--1665} (\bibinfo {year} {1989})}\BibitemShut {NoStop}%
\bibitem [{\citenamefont {Goldstein}\ and\ \citenamefont {andJohn
  Safko}(2001)}]{goldstein}%
  \BibitemOpen
  \bibfield  {author} {\bibinfo {author} {\bibfnamefont {H.}~\bibnamefont
  {Goldstein}}\ and\ \bibinfo {author} {\bibfnamefont {C.~P.~A.}\ \bibnamefont
  {andJohn Safko}},\ }\href@noop {} {\emph {\bibinfo {title} {Classical
  Mechanics}}}\ (\bibinfo  {publisher} {Pearson},\ \bibinfo {year}
  {2001})\BibitemShut {NoStop}%
\bibitem [{\citenamefont {Ou}\ \emph {et~al.}(2014)\citenamefont {Ou},
  \citenamefont {Fatehi}, \citenamefont {Alguire},\ and\ \citenamefont
  {Subotnik}}]{ou:2014:tddft_tda}%
  \BibitemOpen
  \bibfield  {author} {\bibinfo {author} {\bibfnamefont {Q.}~\bibnamefont
  {Ou}}, \bibinfo {author} {\bibfnamefont {S.}~\bibnamefont {Fatehi}}, \bibinfo
  {author} {\bibfnamefont {E.}~\bibnamefont {Alguire}},\ and\ \bibinfo {author}
  {\bibfnamefont {J.~E.}\ \bibnamefont {Subotnik}},\ }\bibfield  {title}
  {\enquote {\bibinfo {title} {Derivative couplings between tddft excited
  states obtained by direct differentiation in the tamm-dancoff
  approximation},}\ }\href@noop {} {\bibfield  {journal} {\bibinfo  {journal}
  {Journal of Chemical Physics}\ }\textbf {\bibinfo {volume} {141}},\ \bibinfo
  {pages} {024114} (\bibinfo {year} {2014})}\BibitemShut {NoStop}%
\bibitem [{\citenamefont {Lengsfield}, \citenamefont {Saxe},\ and\
  \citenamefont {Yarkony}(1984)}]{yarkony:1984:jcp_dercouple}%
  \BibitemOpen
  \bibfield  {author} {\bibinfo {author} {\bibfnamefont {B.~H.}\ \bibnamefont
  {Lengsfield}}, \bibinfo {author} {\bibfnamefont {P.}~\bibnamefont {Saxe}},\
  and\ \bibinfo {author} {\bibfnamefont {D.~R.}\ \bibnamefont {Yarkony}},\
  }\bibfield  {title} {\enquote {\bibinfo {title} {On the evaluation of
  nonadiabatic coupling matrix elements using sa‐mcscf/ci wave functions and
  analytic gradient methods. i},}\ }\href@noop {} {\bibfield  {journal}
  {\bibinfo  {journal} {Journal of Chemical Physics}\ }\textbf {\bibinfo
  {volume} {81}},\ \bibinfo {pages} {4549} (\bibinfo {year}
  {1984})}\BibitemShut {NoStop}%
\bibitem [{\citenamefont {Yarkony}(1984)}]{yarkony:1984:emailme}%
  \BibitemOpen
  \bibfield  {author} {\bibinfo {author} {\bibfnamefont {D.~R.}\ \bibnamefont
  {Yarkony}},\ }\bibfield  {title} {\enquote {\bibinfo {title} {On the reaction
  na($^2$p)+h$_2 \rightarrow $ na($^2$s)+h$_2$ nonadiabatic effects},}\
  }\href@noop {} {\bibfield  {journal} {\bibinfo  {journal} {Journal of
  Chemical Physics}\ }\textbf {\bibinfo {volume} {84}},\ \bibinfo {pages}
  {3206--3211} (\bibinfo {year} {1984})}\BibitemShut {NoStop}%
\bibitem [{\citenamefont {Zhang}\ and\ \citenamefont
  {Herbert}(2014)}]{herbert:2014:jcp_dercouple}%
  \BibitemOpen
  \bibfield  {author} {\bibinfo {author} {\bibfnamefont {X.}~\bibnamefont
  {Zhang}}\ and\ \bibinfo {author} {\bibfnamefont {J.~M.}\ \bibnamefont
  {Herbert}},\ }\bibfield  {title} {\enquote {\bibinfo {title} {Analytic
  derivative couplings for spin-flip configuration interaction singles and
  spin-flip time-dependent density functional theory},}\ }\href@noop {}
  {\bibfield  {journal} {\bibinfo  {journal} {Journal of Chemical Physics}\
  }\textbf {\bibinfo {volume} {141}},\ \bibinfo {eid} {064104} (\bibinfo {year}
  {2014})}\BibitemShut {NoStop}%
\bibitem [{\citenamefont {Li}, \citenamefont {Suo},\ and\ \citenamefont
  {Liu}(2014)}]{liu:2014:dercouple_div}%
  \BibitemOpen
  \bibfield  {author} {\bibinfo {author} {\bibfnamefont {Z.}~\bibnamefont
  {Li}}, \bibinfo {author} {\bibfnamefont {B.}~\bibnamefont {Suo}},\ and\
  \bibinfo {author} {\bibfnamefont {W.}~\bibnamefont {Liu}},\ }\bibfield
  {title} {\enquote {\bibinfo {title} {First order nonadiabatic coupling matrix
  elements between excited states: Implementation and application at the td-dft
  and pp-tda levels},}\ }\href@noop {} {\bibfield  {journal} {\bibinfo
  {journal} {Journal of Chemical Physics}\ }\textbf {\bibinfo {volume} {141}},\
  \bibinfo {eid} {244105} (\bibinfo {year} {2014})}\BibitemShut {NoStop}%
\bibitem [{\citenamefont {Ou}\ \emph {et~al.}(2015)\citenamefont {Ou},
  \citenamefont {Bellchambers}, \citenamefont {Furche},\ and\ \citenamefont
  {Subotnik}}]{ou:2015:dercouple_div}%
  \BibitemOpen
  \bibfield  {author} {\bibinfo {author} {\bibfnamefont {Q.}~\bibnamefont
  {Ou}}, \bibinfo {author} {\bibfnamefont {G.~D.}\ \bibnamefont
  {Bellchambers}}, \bibinfo {author} {\bibfnamefont {F.}~\bibnamefont
  {Furche}},\ and\ \bibinfo {author} {\bibfnamefont {J.~E.}\ \bibnamefont
  {Subotnik}},\ }\bibfield  {title} {\enquote {\bibinfo {title} {First-order
  derivative couplings between excited states from adiabatic tddft response
  theory},}\ }\href@noop {} {\bibfield  {journal} {\bibinfo  {journal} {Journal
  of Chemical Physics}\ }\textbf {\bibinfo {volume} {142}},\ \bibinfo {pages}
  {064114} (\bibinfo {year} {2015})}\BibitemShut {NoStop}%
\bibitem [{\citenamefont {Bates}\ and\ \citenamefont
  {McCarroll}(1958)}]{bates:1958:etf}%
  \BibitemOpen
  \bibfield  {author} {\bibinfo {author} {\bibfnamefont {D.~R.}\ \bibnamefont
  {Bates}}\ and\ \bibinfo {author} {\bibfnamefont {R.}~\bibnamefont
  {McCarroll}},\ }\bibfield  {title} {\enquote {\bibinfo {title} {Electron
  capture in slow collisions},}\ }\href@noop {} {\bibfield  {journal} {\bibinfo
   {journal} {Proc. R. Soc. A}\ }\textbf {\bibinfo {volume} {245}},\ \bibinfo
  {pages} {175} (\bibinfo {year} {1958})}\BibitemShut {NoStop}%
\bibitem [{\citenamefont {Schneiderman}\ and\ \citenamefont
  {Russek}(1969)}]{schneiderman:1969:pr:etf}%
  \BibitemOpen
  \bibfield  {author} {\bibinfo {author} {\bibfnamefont {S.~B.}\ \bibnamefont
  {Schneiderman}}\ and\ \bibinfo {author} {\bibfnamefont {A.}~\bibnamefont
  {Russek}},\ }\bibfield  {title} {\enquote {\bibinfo {title}
  {Velocity-dependent orbitals in proton-on-hydrogen-atom collisions},}\ }\href
  {https://doi.org/10.1103/PhysRev.181.311} {\bibfield  {journal} {\bibinfo
  {journal} {Physical Review}\ }\textbf {\bibinfo {volume} {181}},\ \bibinfo
  {pages} {311--321} (\bibinfo {year} {1969})}\BibitemShut {NoStop}%
\bibitem [{\citenamefont {Thorson}\ and\ \citenamefont
  {Delos}(1978)}]{delos:1978:pra:etf}%
  \BibitemOpen
  \bibfield  {author} {\bibinfo {author} {\bibfnamefont {W.~R.}\ \bibnamefont
  {Thorson}}\ and\ \bibinfo {author} {\bibfnamefont {J.~B.}\ \bibnamefont
  {Delos}},\ }\bibfield  {title} {\enquote {\bibinfo {title} {Theory of
  near-adiabatic collisions. i. electron translation factor method},}\ }\href
  {https://doi.org/10.1103/PhysRevA.18.117} {\bibfield  {journal} {\bibinfo
  {journal} {Physical Review A}\ }\textbf {\bibinfo {volume} {18}},\ \bibinfo
  {pages} {117--134} (\bibinfo {year} {1978})}\BibitemShut {NoStop}%
\bibitem [{\citenamefont {Delos}(1981)}]{delos:1981:rmp}%
  \BibitemOpen
  \bibfield  {author} {\bibinfo {author} {\bibfnamefont {J.~B.}\ \bibnamefont
  {Delos}},\ }\bibfield  {title} {\enquote {\bibinfo {title} {Theory of
  electronic transitions in slow atomic collisions},}\ }\href@noop {}
  {\bibfield  {journal} {\bibinfo  {journal} {Reviews of Modern Physics}\
  }\textbf {\bibinfo {volume} {53}},\ \bibinfo {pages} {287--357} (\bibinfo
  {year} {1981})}\BibitemShut {NoStop}%
\bibitem [{\citenamefont {Winter}(1982)}]{winter:1982:pra:etf}%
  \BibitemOpen
  \bibfield  {author} {\bibinfo {author} {\bibfnamefont {T.~G.}\ \bibnamefont
  {Winter}},\ }\bibfield  {title} {\enquote {\bibinfo {title} {Electron
  transfer in $p\ensuremath{-}{\mathrm{he}}^{+}$ and ${\mathrm{he}}^{2+}$-h
  collisions using a sturmian basis},}\ }\href
  {https://doi.org/10.1103/PhysRevA.25.697} {\bibfield  {journal} {\bibinfo
  {journal} {Physical Review A}\ }\textbf {\bibinfo {volume} {25}},\ \bibinfo
  {pages} {697--712} (\bibinfo {year} {1982})}\BibitemShut {NoStop}%
\bibitem [{\citenamefont {Errea}\ \emph {et~al.}(1994)\citenamefont {Errea},
  \citenamefont {Harel}, \citenamefont {Jouini}, \citenamefont {Mendez},
  \citenamefont {Pons},\ and\ \citenamefont {Riera}}]{errea:1994:etf}%
  \BibitemOpen
  \bibfield  {author} {\bibinfo {author} {\bibfnamefont {L.~F.}\ \bibnamefont
  {Errea}}, \bibinfo {author} {\bibfnamefont {C.}~\bibnamefont {Harel}},
  \bibinfo {author} {\bibfnamefont {H.}~\bibnamefont {Jouini}}, \bibinfo
  {author} {\bibfnamefont {L.}~\bibnamefont {Mendez}}, \bibinfo {author}
  {\bibfnamefont {B.}~\bibnamefont {Pons}},\ and\ \bibinfo {author}
  {\bibfnamefont {A.}~\bibnamefont {Riera}},\ }\bibfield  {title} {\enquote
  {\bibinfo {title} {Common translation factor method},}\ }\href
  {https://doi.org/10.1088/0953-4075/27/16/010} {\bibfield  {journal} {\bibinfo
   {journal} {Journal of Physics B: Atomic, Molecular and Optical Physics}\
  }\textbf {\bibinfo {volume} {27}},\ \bibinfo {pages} {3603} (\bibinfo {year}
  {1994})}\BibitemShut {NoStop}%
\bibitem [{\citenamefont {Deumens}\ \emph {et~al.}(1994)\citenamefont
  {Deumens}, \citenamefont {Diz}, \citenamefont {Longo},\ and\ \citenamefont
  {\"Ohrn}}]{ohrn:1994:rmp:etf}%
  \BibitemOpen
  \bibfield  {author} {\bibinfo {author} {\bibfnamefont {E.}~\bibnamefont
  {Deumens}}, \bibinfo {author} {\bibfnamefont {A.}~\bibnamefont {Diz}},
  \bibinfo {author} {\bibfnamefont {R.}~\bibnamefont {Longo}},\ and\ \bibinfo
  {author} {\bibfnamefont {Y.}~\bibnamefont {\"Ohrn}},\ }\bibfield  {title}
  {\enquote {\bibinfo {title} {Time-dependent theoretical treatments of the
  dynamics of electrons and nuclei in molecular systems},}\ }\href
  {https://doi.org/10.1103/RevModPhys.66.917} {\bibfield  {journal} {\bibinfo
  {journal} {Reviews of Modern Physics}\ }\textbf {\bibinfo {volume} {66}},\
  \bibinfo {pages} {917--983} (\bibinfo {year} {1994})}\BibitemShut {NoStop}%
\bibitem [{\citenamefont {Illescas}\ and\ \citenamefont
  {Riera}(1998)}]{riera:1998:PRL}%
  \BibitemOpen
  \bibfield  {author} {\bibinfo {author} {\bibfnamefont {C.}~\bibnamefont
  {Illescas}}\ and\ \bibinfo {author} {\bibfnamefont {A.}~\bibnamefont
  {Riera}},\ }\bibfield  {title} {\enquote {\bibinfo {title} {Classical outlook
  on the electron translation factor problem},}\ }\href
  {https://doi.org/10.1103/PhysRevLett.80.3029} {\bibfield  {journal} {\bibinfo
   {journal} {Physical Review Letters}\ }\textbf {\bibinfo {volume} {80}},\
  \bibinfo {pages} {3029--3032} (\bibinfo {year} {1998})}\BibitemShut {NoStop}%
\bibitem [{\citenamefont {Littlejohn}, \citenamefont {Rawlinson},\ and\
  \citenamefont {Subotnik}(2023)}]{littlejohn:2023:ang}%
  \BibitemOpen
  \bibfield  {author} {\bibinfo {author} {\bibfnamefont {R.}~\bibnamefont
  {Littlejohn}}, \bibinfo {author} {\bibfnamefont {J.}~\bibnamefont
  {Rawlinson}},\ and\ \bibinfo {author} {\bibfnamefont {J.}~\bibnamefont
  {Subotnik}},\ }\bibfield  {title} {\enquote {\bibinfo {title} {Representation
  and conservation of angular momentum in the born--oppenheimer theory of
  polyatomic molecules},}\ }\href@noop {} {\bibfield  {journal} {\bibinfo
  {journal} {The Journal of Chemical Physics}\ }\textbf {\bibinfo {volume}
  {158}},\ \bibinfo {pages} {104302} (\bibinfo {year} {2023})}\BibitemShut
  {NoStop}%
\bibitem [{Note1()}]{Note1}%
  \BibitemOpen
  \bibinfo {note} {The fact that Eq. \ref {finaleq-op} is relevant only for
  rotations becomes obvious when one considers the fact that $\protect \genfrac
  {}{}{}{}{\partial S^{op}_{\mu \nu }}{\partial R_{A \beta }} = \protect
  \genfrac {}{}{}{}{\partial \pi ^{op}_{\mu \nu \lambda \sigma }}{\partial R_{A
  \beta }} = 0$. Thus, if Eq. \ref {explbasis} were relevant more generally,
  there would be no Pulay terms or two-electron derivative terms.}\BibitemShut
  {Stop}%
\bibitem [{\citenamefont {Epifanovsky}\ \emph {et~al.}(2021)\citenamefont
  {Epifanovsky}, \citenamefont {Gilbert}, \citenamefont {Feng}, \citenamefont
  {Lee}, \citenamefont {Mao}, \citenamefont {Mardirossian}, \citenamefont
  {Pokhilko}, \citenamefont {White}, \citenamefont {Coons}, \citenamefont
  {Dempwolff} \emph {et~al.}}]{qchem6}%
  \BibitemOpen
  \bibfield  {author} {\bibinfo {author} {\bibfnamefont {E.}~\bibnamefont
  {Epifanovsky}}, \bibinfo {author} {\bibfnamefont {A.~T.}\ \bibnamefont
  {Gilbert}}, \bibinfo {author} {\bibfnamefont {X.}~\bibnamefont {Feng}},
  \bibinfo {author} {\bibfnamefont {J.}~\bibnamefont {Lee}}, \bibinfo {author}
  {\bibfnamefont {Y.}~\bibnamefont {Mao}}, \bibinfo {author} {\bibfnamefont
  {N.}~\bibnamefont {Mardirossian}}, \bibinfo {author} {\bibfnamefont
  {P.}~\bibnamefont {Pokhilko}}, \bibinfo {author} {\bibfnamefont {A.~F.}\
  \bibnamefont {White}}, \bibinfo {author} {\bibfnamefont {M.~P.}\ \bibnamefont
  {Coons}}, \bibinfo {author} {\bibfnamefont {A.~L.}\ \bibnamefont
  {Dempwolff}}, \emph {et~al.},\ }\bibfield  {title} {\enquote {\bibinfo
  {title} {Software for the frontiers of quantum chemistry: An overview of
  developments in the q-chem 5 package},}\ }\href@noop {} {\bibfield  {journal}
  {\bibinfo  {journal} {Journal of Chemical Physics}\ }\textbf {\bibinfo
  {volume} {155}},\ \bibinfo {pages} {084801} (\bibinfo {year}
  {2021})}\BibitemShut {NoStop}%
\bibitem [{\citenamefont {Hammes-Schiffer}\ and\ \citenamefont
  {Tully}(1994)}]{shs:1994:protons}%
  \BibitemOpen
  \bibfield  {author} {\bibinfo {author} {\bibfnamefont {S.}~\bibnamefont
  {Hammes-Schiffer}}\ and\ \bibinfo {author} {\bibfnamefont {J.}~\bibnamefont
  {Tully}},\ }\bibfield  {title} {\enquote {\bibinfo {title} {Proton transfer
  in solution: Molecular dynamics with quantum transitions},}\ }\href@noop {}
  {\bibfield  {journal} {\bibinfo  {journal} {Journal of Chemical Physics}\
  }\textbf {\bibinfo {volume} {101}},\ \bibinfo {pages} {4657--4667} (\bibinfo
  {year} {1994})}\BibitemShut {NoStop}%
\bibitem [{\citenamefont {Meek}\ and\ \citenamefont {Levine}(2014)}]{Meek2014}%
  \BibitemOpen
  \bibfield  {author} {\bibinfo {author} {\bibfnamefont {G.~A.}\ \bibnamefont
  {Meek}}\ and\ \bibinfo {author} {\bibfnamefont {B.~G.}\ \bibnamefont
  {Levine}},\ }\bibfield  {title} {\enquote {\bibinfo {title} {Evaluation of
  the time-derivative coupling for accurate electronic state transition
  probabilities from numerical simulations},}\ }\href@noop {} {\bibfield
  {journal} {\bibinfo  {journal} {The journal of physical chemistry letters}\
  }\textbf {\bibinfo {volume} {5}},\ \bibinfo {pages} {2351--2356} (\bibinfo
  {year} {2014})}\BibitemShut {NoStop}%
\bibitem [{\citenamefont {Jain}, \citenamefont {Alguire},\ and\ \citenamefont
  {Subotnik}(2016)}]{jain:2016:fast}%
  \BibitemOpen
  \bibfield  {author} {\bibinfo {author} {\bibfnamefont {A.}~\bibnamefont
  {Jain}}, \bibinfo {author} {\bibfnamefont {E.}~\bibnamefont {Alguire}},\ and\
  \bibinfo {author} {\bibfnamefont {J.~E.}\ \bibnamefont {Subotnik}},\
  }\bibfield  {title} {\enquote {\bibinfo {title} {An efficient, augmented
  surface hopping algorithm that includes decoherence for use in large-scale
  simulations},}\ }\href {https://doi.org/10.1021/acs.jctc.6b00673} {\bibfield
  {journal} {\bibinfo  {journal} {Journal of Chemical Theory and Computation}\
  }\textbf {\bibinfo {volume} {12}},\ \bibinfo {pages} {5256--5268} (\bibinfo
  {year} {2016})},\ \bibinfo {note} {pMID: 27715036},\ \Eprint
  {https://arxiv.org/abs/http://dx.doi.org/10.1021/acs.jctc.6b00673}
  {http://dx.doi.org/10.1021/acs.jctc.6b00673} \BibitemShut {NoStop}%
\bibitem [{\citenamefont {Bian}\ \emph {et~al.}(2023)\citenamefont {Bian},
  \citenamefont {Tao}, \citenamefont {Wu}, \citenamefont {Rawlinson},
  \citenamefont {Littlejohn},\ and\ \citenamefont
  {Subotnik}}]{bian:2023:angcons}%
  \BibitemOpen
  \bibfield  {author} {\bibinfo {author} {\bibfnamefont {X.}~\bibnamefont
  {Bian}}, \bibinfo {author} {\bibfnamefont {Z.}~\bibnamefont {Tao}}, \bibinfo
  {author} {\bibfnamefont {Y.}~\bibnamefont {Wu}}, \bibinfo {author}
  {\bibfnamefont {J.}~\bibnamefont {Rawlinson}}, \bibinfo {author}
  {\bibfnamefont {R.~G.}\ \bibnamefont {Littlejohn}},\ and\ \bibinfo {author}
  {\bibfnamefont {J.~E.}\ \bibnamefont {Subotnik}},\ }\bibfield  {title}
  {\enquote {\bibinfo {title} {Total angular momentum conservation in ab initio
  born-oppenheimer molecular dynamics},}\ }\href@noop {} {\bibfield  {journal}
  {\bibinfo  {journal} {arXiv preprint arXiv:2303.13787}\ } (\bibinfo {year}
  {2023})}\BibitemShut {NoStop}%
\bibitem [{\citenamefont {Naaman}\ and\ \citenamefont
  {Waldeck}(2012)}]{naaman:2012:jpcl}%
  \BibitemOpen
  \bibfield  {author} {\bibinfo {author} {\bibfnamefont {R.}~\bibnamefont
  {Naaman}}\ and\ \bibinfo {author} {\bibfnamefont {D.~H.}\ \bibnamefont
  {Waldeck}},\ }\bibfield  {title} {\enquote {\bibinfo {title} {Chiral-induced
  spin selectivity effect},}\ }\href@noop {} {\bibfield  {journal} {\bibinfo
  {journal} {Journal of Physical Chemistry Letters}\ }\textbf {\bibinfo
  {volume} {3}},\ \bibinfo {pages} {2178--2187} (\bibinfo {year}
  {2012})}\BibitemShut {NoStop}%
\bibitem [{\citenamefont {Naaman}, \citenamefont {Paltiel},\ and\ \citenamefont
  {Waldeck}(2019)}]{naaman:2019:natrev}%
  \BibitemOpen
  \bibfield  {author} {\bibinfo {author} {\bibfnamefont {R.}~\bibnamefont
  {Naaman}}, \bibinfo {author} {\bibfnamefont {Y.}~\bibnamefont {Paltiel}},\
  and\ \bibinfo {author} {\bibfnamefont {D.~H.}\ \bibnamefont {Waldeck}},\
  }\bibfield  {title} {\enquote {\bibinfo {title} {Chiral molecules and the
  electron spin},}\ }\href@noop {} {\bibfield  {journal} {\bibinfo  {journal}
  {Nature Reviews Chemistry}\ }\textbf {\bibinfo {volume} {3}},\ \bibinfo
  {pages} {250--260} (\bibinfo {year} {2019})}\BibitemShut {NoStop}%
\bibitem [{\citenamefont {Das}\ \emph {et~al.}(2022)\citenamefont {Das},
  \citenamefont {Tassinari}, \citenamefont {Naaman},\ and\ \citenamefont
  {Fransson}}]{das:2022:cisstemp}%
  \BibitemOpen
  \bibfield  {author} {\bibinfo {author} {\bibfnamefont {T.~K.}\ \bibnamefont
  {Das}}, \bibinfo {author} {\bibfnamefont {F.}~\bibnamefont {Tassinari}},
  \bibinfo {author} {\bibfnamefont {R.}~\bibnamefont {Naaman}},\ and\ \bibinfo
  {author} {\bibfnamefont {J.}~\bibnamefont {Fransson}},\ }\bibfield  {title}
  {\enquote {\bibinfo {title} {Temperature-dependent chiral-induced spin
  selectivity effect: Experiments and theory},}\ }\href@noop {} {\bibfield
  {journal} {\bibinfo  {journal} {The Journal of Physical Chemistry C}\
  }\textbf {\bibinfo {volume} {126}},\ \bibinfo {pages} {3257--3264} (\bibinfo
  {year} {2022})}\BibitemShut {NoStop}%
\bibitem [{\citenamefont {Cox}\ and\ \citenamefont
  {Sivia}(1997)}]{cox:1997:hyperfine}%
  \BibitemOpen
  \bibfield  {author} {\bibinfo {author} {\bibfnamefont {S.}~\bibnamefont
  {Cox}}\ and\ \bibinfo {author} {\bibfnamefont {D.}~\bibnamefont {Sivia}},\
  }\bibfield  {title} {\enquote {\bibinfo {title} {Spin-lattice relaxation in
  hyperfine-coupled systems: applications to interstitial diffusion and
  molecular dynamics},}\ }\href@noop {} {\bibfield  {journal} {\bibinfo
  {journal} {Applied Magnetic Resonance}\ }\textbf {\bibinfo {volume} {12}},\
  \bibinfo {pages} {213--226} (\bibinfo {year} {1997})}\BibitemShut {NoStop}%
\bibitem [{\citenamefont {Zheng}, \citenamefont {Zheng},\ and\ \citenamefont
  {Zhao}(2022)}]{zheng:2022:shspinlatt}%
  \BibitemOpen
  \bibfield  {author} {\bibinfo {author} {\bibfnamefont {Z.}~\bibnamefont
  {Zheng}}, \bibinfo {author} {\bibfnamefont {Q.}~\bibnamefont {Zheng}},\ and\
  \bibinfo {author} {\bibfnamefont {J.}~\bibnamefont {Zhao}},\ }\bibfield
  {title} {\enquote {\bibinfo {title} {Spin-orbit coupling induced
  demagnetization in ni: Ab initio nonadiabatic molecular dynamics
  perspective},}\ }\href@noop {} {\bibfield  {journal} {\bibinfo  {journal}
  {Physical Review B}\ }\textbf {\bibinfo {volume} {105}},\ \bibinfo {pages}
  {085142} (\bibinfo {year} {2022})}\BibitemShut {NoStop}%
\bibitem [{\citenamefont {Wu}\ \emph {et~al.}(2022)\citenamefont {Wu},
  \citenamefont {Bian}, \citenamefont {Rawlinson}, \citenamefont {Littlejohn},\
  and\ \citenamefont {Subotnik}}]{wu:2022:pssh}%
  \BibitemOpen
  \bibfield  {author} {\bibinfo {author} {\bibfnamefont {Y.}~\bibnamefont
  {Wu}}, \bibinfo {author} {\bibfnamefont {X.}~\bibnamefont {Bian}}, \bibinfo
  {author} {\bibfnamefont {J.~I.}\ \bibnamefont {Rawlinson}}, \bibinfo {author}
  {\bibfnamefont {R.~G.}\ \bibnamefont {Littlejohn}},\ and\ \bibinfo {author}
  {\bibfnamefont {J.~E.}\ \bibnamefont {Subotnik}},\ }\bibfield  {title}
  {\enquote {\bibinfo {title} {A phase-space semiclassical approach for
  modeling nonadiabatic nuclear dynamics with electronic spin},}\ }\href@noop
  {} {\bibfield  {journal} {\bibinfo  {journal} {The Journal of chemical
  physics}\ }\textbf {\bibinfo {volume} {157}},\ \bibinfo {pages} {011101}
  (\bibinfo {year} {2022})}\BibitemShut {NoStop}%
\bibitem [{\citenamefont {Bian}\ \emph {et~al.}(2022)\citenamefont {Bian},
  \citenamefont {Wu}, \citenamefont {Rawlinson}, \citenamefont {Littlejohn},\
  and\ \citenamefont {Subotnik}}]{bian:2022:pssh}%
  \BibitemOpen
  \bibfield  {author} {\bibinfo {author} {\bibfnamefont {X.}~\bibnamefont
  {Bian}}, \bibinfo {author} {\bibfnamefont {Y.}~\bibnamefont {Wu}}, \bibinfo
  {author} {\bibfnamefont {J.}~\bibnamefont {Rawlinson}}, \bibinfo {author}
  {\bibfnamefont {R.~G.}\ \bibnamefont {Littlejohn}},\ and\ \bibinfo {author}
  {\bibfnamefont {J.~E.}\ \bibnamefont {Subotnik}},\ }\bibfield  {title}
  {\enquote {\bibinfo {title} {Modeling spin-dependent nonadiabatic dynamics
  with electronic degeneracy: a phase-space surface-hopping method},}\
  }\href@noop {} {\bibfield  {journal} {\bibinfo  {journal} {The Journal of
  Physical Chemistry Letters}\ }\textbf {\bibinfo {volume} {13}},\ \bibinfo
  {pages} {7398--7404} (\bibinfo {year} {2022})}\BibitemShut {NoStop}%
\end{thebibliography}%

\end{document}